\documentclass[times, 10pt]{article}
\usepackage{times}
\usepackage{graphicx}
\usepackage[utf8]{inputenc}
\usepackage[T1]{fontenc}
\usepackage{amsmath,amsfonts,amssymb}
\usepackage{url}
\usepackage{multicol}
\usepackage[dvipsnames]{xcolor}
\usepackage{hyperref}
\usepackage{cleveref}

\newcommand\myshade{85}
\colorlet{mylinkcolor}{purple}
\colorlet{mycitecolor}{YellowOrange}
\colorlet{myurlcolor}{NavyBlue}

\hypersetup{
  linkcolor  = mylinkcolor!\myshade!black,
  citecolor  = mycitecolor!\myshade!black,
  urlcolor   = myurlcolor!\myshade!black,
  colorlinks = true,
}

\renewcommand{\vec}[1]{\mathbf{#1}}
\newcommand{\vu}{\vec{u}}
\newcommand{\nor}{\vec{n}}
\newcommand{\vx}{\vec{x}}
\newcommand{\vX}{\vec{X}}

\newcommand{\vF}{\vec{F}}
\newcommand{\Ident}{\mathbf{1}}
\newcommand{\vxi}{\boldsymbol\xi}
\newcommand{\vsigma}{\boldsymbol\sigma}
\newcommand{\vbeta}{\boldsymbol\beta}
\newcommand{\vtau}{\boldsymbol\tau}
\newcommand{\vchi}{\boldsymbol\chi}
\newcommand{\Rey}{\textit{Re}}
\newcommand{\trans}{\mathsf{T}}
\DeclareMathOperator{\Tr}{tr}

\newcommand\xhat{\vec{e}_X}

\newcommand\Xhat{\tilde{\vec{e}}_X}

\newcommand\Zhat{\tilde{\vec{e}}_Z}
\newcommand\ihat{\vec{e}_I}
\newcommand\Ihat{\tilde{\vec{e}}_I}

\newcommand\vI{\vec 1}

\usepackage{authblk}
\usepackage[colorinlistoftodos]{todonotes}

\definecolor{purp}{rgb}{0.4,0.2,0.8}

\title{Eulerian simulation of complex suspensions and biolocomotion in three dimensions}
\author[a]{Yuexia L. Lin}
\author[a]{Nicholas J. Derr}
\author[a,b]{Chris H. Rycroft\thanks{Corresponding author. E-mail: chr@seas.harvard.edu}}

\affil[a]{John A. Paulson School of Engineering and Applied Sciences, Harvard University, 29 Oxford Street, Cambridge, MA 02138}
\affil[b]{Mathematics Group, Lawrence Berkeley National Laboratory, 1 Cyclotron Road, Berkeley, CA 94720}
\date{}

\begin{document}
\maketitle

\thispagestyle{plain}

\begin{abstract}
We present a numerical method specifically designed for simulating three-dimensional fluid--structure interaction (FSI) problems based on the reference map technique (RMT). The RMT is a fully Eulerian FSI numerical method that allows fluids and large-deformation elastic solids to be represented on a single fixed computational grid. This eliminates the need for meshing complex geometries typical in other FSI approaches, and greatly simplifies the coupling between fluid and solids. We develop the first three-dimensional implementation of the RMT, parallelized using the distributed memory paradigm, to simulate incompressible FSI with neo-Hookean solids. As part of our new method, we develop a new field extrapolation scheme that works efficiently in parallel. Through representative examples, we demonstrate the method's accuracy and convergence, as well as its suitability in investigating many-body and active systems.
The examples include settling of a mixture of heavy and buoyant soft ellipsoids, lid-driven cavity flow containing a soft sphere, and swimmers actuated via active stress.
\end{abstract}

\begin{multicols}{2}
\section*{Introduction}

Fluid--structure interactions (FSI) are at the heart of many important physical and biological problems,
including flexible structures in flow \cite{luhar11, bukowicki19},
blood circulation in the heart \cite{peskin72, griffith09},
animal locomotion \cite{lucas14,lim2019}, and cilia motion \cite{kanso17,kanso2018}.
The couplings between fluid and immersed solids give rise to complex nonlinear dynamics dependent on geometry and boundary conditions, material constitutive relations, and collective interactions among the solid objects.
Thus analytical solutions are rare and limited to simplified settings in reduced dimensions, and numerical methods for FSI have become indispensable for understand these problems.

In designing numerical methods for fluids and solids, Eulerian and Lagrangian perspectives are the more convenient choices, respectively, due to the differences in constitutive responses.
Bridging between the two perspectives is a classic dilemma in developing numerical methods for FSI.
Various frameworks have been proposed to resolve this dilemma.
The immersed boundary method \cite{peskin72, peskin02} and the family of immersed methods that it inspires \cite{griffith09, fai18, griffith20} solve fluid on a fixed mesh, use Lagrangian points to represent the solid, and employ a coupling scheme between the two.
Arbitrary Lagrangian--Eulerian methods use moving nodes for both phases, and reposition the nodes to maintain mesh quality \cite{hirt74, rugonyi01}.

There are also fully Eulerian methods, for which two types of formulations exist.
The hypoelasticity formulation \cite{truesdell55, udaykumar03, rycroft12} uses linear elasticity, whereas the hyperelasticity formulation employs a general large-deformation description in the solids.
To compute solid stresses in hyperelasticity, a variety of quantities have been used, such as level-set functions defining the fluid--solid interface \cite{maitre09}, the deformation gradient tensor \cite{liu01}, the deformation tensors \cite{sugiyama11}, and the solid displacements in the undeformed configuration \cite{dunne06, richter13, wick13}.

The reference map technique (RMT) is a fully Eulerian approach that uses the solid's reference coordinates as the primary simulation variables. The RMT has many favorable properties, and can simulate compressible fluid and solids \cite{kamrin_thesis,kamrin12}, handle contact between multiple solids \cite{valkov15}, and simulate incompressible fluid and solids with complex geometries and actuation \cite{rycroft20}. The RMT makes use of the level-set method \cite{osher88, sethian99} for tracking the fluid--solid interfaces, and all prior versions have employed level-set reinitialization using the fast marching method (FMM) \cite{sethian99,rycroft12}. Furthermore the FMM is used to extrapolate field values near the fluid--solid interface, which is a necessary part of the RMT.

All prior papers have demonstrated the method in two dimensions, yet a major strength of the fully Eulerian approach is the ability to avoid computational meshing of complex geometries, which becomes a compelling advantage in three dimensions (3D). Here, we provide the first 3D implementation of the RMT, and use it to simulate several scenarios that would be difficult do with existing numerical approaches. Inspired by recent work on fluid-filled soft granular packings \cite{macminn15}, we simulate a suspension of soft flexible particles, half of which are lighter and half of which are heavier than the fluid. There has also been substantial recent interest in swimming and biolocomotion, yet reduced-dimension models of the swimmer \cite{tytell2010,thomases2014,olson2020} are often the state of the art. We demonstrate a model of a swimming organism using full volumetric actuation, creating a simulation tool that can explore a broader range of swimming modalities.

3D calculations are considerably more computationally challenging than two-dimensional (2D) ones, and we therefore present the first complete parallel implementation of the method using domain decomposition. Since the RMT uses a single fixed regular grid for fluid and solid, it is well-suited to parallelization, and different processors can each handle a rectangular subdomain. In addition, the regular grid structure makes contact among solid bodies easy to detect and allows for efficient memory storage.
One challenge that we faced is that the FMM is not well-suited for parallelization, since it updates field values sequentially. We therefore a developed new method for extrapolation that can be done in parallel. Our new method removes the need for level-set reinitialization and fast marching methods, and the RMT implementation is therefore simplified from all prior approaches.

\section*{Theory and numerical method}

\subsection*{Hyperelastic formulation}

             \begin{figure*}[ht]
                \centering
                \includegraphics[width=8.7cm, keepaspectratio]{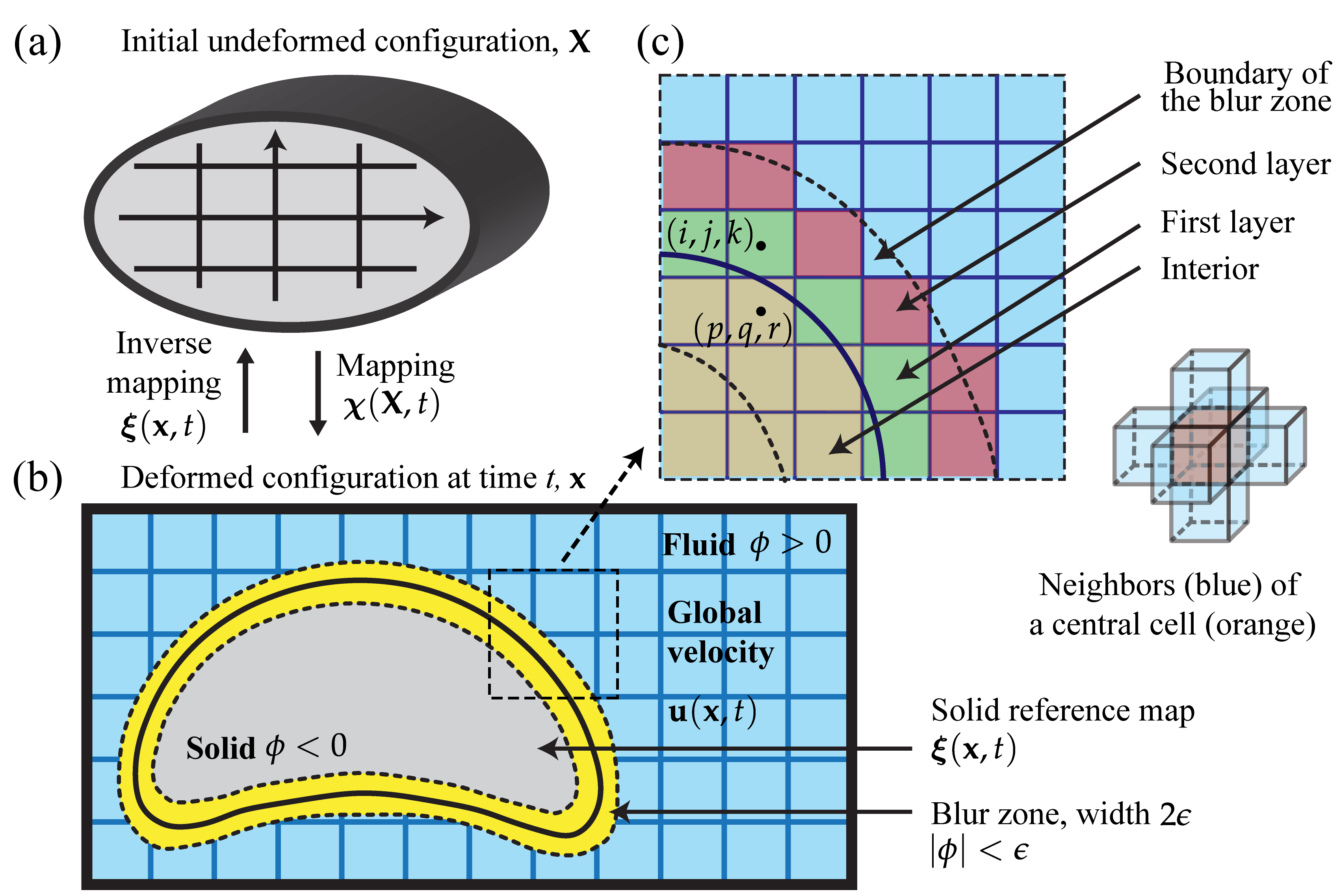}
                 \caption{
	        (a) An initially undeformed solid with reference configuration $\vX$ undergoes a
                time-dependent mapping $\vchi(\vX,t)$ to its current configuration at time
                $t$.
                (b) A level-set function $\phi(\vx,t)$ distinguishes the two phases which share
                a global velocity $\vu(\vx,t)$.
                A blur zone (yellow), defined by $|\phi|<\epsilon$, is
                used to transition between phases.
                (c) The order in which $\vxi(\vx, t)$ is extrapolated is defined by layers.
                The $1^\text{st}$ layer (green), e.g.~cell $(i, j, k)$, are non-interior orthogonal neighbors to the interior cells (yellow), e.g.~cell $(p, q, r)$.
                Subsequent layers, e.g.~$2^\text{nd}$ layers (red), are constructed in the same way,
                until the blur zone is filled or a physical boundary is reached.
                2D schematics are shown for clarity.
                \label{fig:overview}}
            \end{figure*}

In the hyperelasticity framework \cite{gurtin10}, a time-dependent mapping function $ \vchi (\vX, t)$
is introduced to determine how
the undeformed configuration, $\vX$,
is transformed to its current physical configuration, $\vx$, i.e.~$\vx =  \vchi(\vX, t)$ (Fig.~\ref{fig:overview}(a,~b)).
The deformation gradient is defined as $\vF = \tfrac{\partial  \vchi}{\partial \vX} = \tfrac{\partial \vx}{\partial \vX}$.
A constitutive relation $ \vsigma_s (\vF)$ defines the Cauchy stress response in the solid material.
The solid momentum balance equation in rate form is
\begin{equation}
\label{eq:solid_mmt}
\rho_s \left (\frac{\partial \vu}{\partial t}
 + ( \vu \cdot \nabla ) \vu \right) = \nabla \cdot \vsigma_s
 \end{equation}
where $\rho_s$ is the solid density and $\vu$ is the solid velocity.
For an incompressible material $\det \vF = 1$ and thus solid density is unaffected by the deformation.
To proceed, we assume that $\vchi$ is a sufficiently smooth function of $\vX$ and $t$ and define its inverse as the reference map, $ \vX = \vxi (\vx, t) = \vchi^{-1} (\vx, t)$.
The deformation gradient tensor becomes
\begin{equation}
\label{eq:def_grad}
\vF = \left(\tfrac{\partial \vxi}{\partial \vx}\right)^{-1} = \left( \mathbf \nabla_{\vx}  \vxi \right)^{-1}
\end{equation}
where $\mathbf \nabla_{\vx}$ is the gradient operator in physical space.
The reference configuration is constant, therefore $\dot{\vxi} (\vx,t) = \vec{0}$, i.e.,
\begin{equation}
\frac{\partial  \vxi}{\partial t} + (\vu \cdot \nabla)  \vxi = \vec{0}.
\label{eq:xi_adv}
\end{equation}
We discuss coupling fluid and solid phases and imposing the incompressibility constraint next.

%
\subsection*{Blurred interface method and monolithic governing equations}
Consider a domain $\Omega$ containing $n$ immersed solid objects covering subdomains $\Omega_1,\ldots,\Omega_n$.
Denote the fluid domain as $\Omega_f$ and the solid domain $\Omega_s = \cup_{k=1}^{n} \Omega_k$ for $k=1,2,\ldots,n$, so that $\Omega = \Omega_s \cup \Omega_f$.
The fluid--solid interface (hereinafter the interface) is denoted as $\partial \Omega_s = \cup_{k=1}^{n} \partial \Omega_k$ for $k=1,2,\ldots,n$.
To avoid excessive distortion, the reference map $\vxi$ is only defined and evolved within $\Omega_s$.
The solution to Eq.~\eqref{eq:xi_adv} is the union of solutions within each $\Omega_k$.
The velocity $\vu$ is defined as a global variable spanning $\Omega$. The incompressibility constraint implies that
\begin{equation}
\label{eq:incomp}
\nabla \cdot \vu = 0
\end{equation}
in $\Omega$. To enforce the constraint, a global pressure field is used as a Lagrange multiplier.
Thus, we need only consider the deviatoric part of the stress tensors.
For the fluid, we consider the deviatoric stress of a viscous Newtonian fluid,
$ \vtau_f = \mu \left(\nabla \vu + \nabla\vu^\trans \right)$,
where $\mu$ is the fluid dynamic viscosity.
The deviatoric solid stress is defined as $ \vtau_s = \vsigma_s - \tfrac13 \Tr ( \vsigma_s)\Ident$.

In a physical system, there may be discontinuities in velocity, density, stresses, and forces across the interface.
In our method, we consider a continuous velocity field, which naturally corresponds to a no-slip boundary condition at $\partial \Omega_s$.
Next, we discuss the blurred interface method that ensures the traction-matching condition, i.e.~$\vsigma_s \cdot \nor  = \vsigma_f \cdot \nor$,
is satisfied at the interface, and creates a smooth transition of field values across the interface.

The basic RMT equations work with a variety of interfacial coupling procedures, including sharp and blurred interface methods \cite{kamrin12, valkov15, rycroft20}.
In this work we focus on a blurred interface method because it has
several advantages over a sharp interface method,
e.g.,~it is more stable to interfacial perturbations, more amenable to simulating immersed solid--solid contact, and easier to implement \cite{valkov15, rycroft20}.
Continuing with the blurred interface method, we make use of a blur zone of width $2\epsilon$ across the interface.
The width of the blur zone $2\epsilon$ scales with the grid spacing so that as grid spacing approaches zero, a sharp interface representation is recovered.
Without loss of generality, we consider a single solid object with interface $\partial \Omega_{s}$ defined by the zero contour of a signed-distance function in the undeformed configuration, $\phi_0(\vX)$.
Throughout the simulation, we define the time-dependent level-set function $\phi(\vx,t) = \phi_0(\vxi(\vx,t))$ (Fig.~\ref{fig:overview}(b)).

In the blur zone, quantities that may have jump across the interface, e.g.~$\rho$, $\vtau$ and $\mathbf b$, are blended to smoothly vary between the fluid and the solid phases \cite{sussman94,sussman99,yu03}.
Let $Q$ denote a scalar, or a component of a vector or a tensorial quantity in $\Omega$.
Consider $Q_f$ in the fluid domain $\Omega_f$, and $Q_s$ in the solid domain $\Omega_k$.
We blend $Q_f$ and $Q_s$ such that $Q$ transitions smoothly between $\Omega_f$ and $\Omega_k$,
\begin{equation}
\label{eq:blend}
Q = Q_s + H_\epsilon(\phi) (Q_f - Q_s)
\end{equation}
where $\phi$ is the value of the level-set function defining the boundary of $\Omega_k$, and $H_\epsilon(\phi)$ is a smoothed Heaviside function,
\begin{equation}
  \label{eq:sm_heavi}
  H_\epsilon(\phi) =
  \left\{
  \begin{array}{ll}
    0 & \qquad \text{if $\phi\le-\epsilon$,} \\
    \frac{1}{2} (1+\frac{\phi}{\epsilon}+\frac{1}{\pi} \sin \frac{\pi \phi}{\epsilon}) & \qquad \text{if $|\phi|<\epsilon$,} \\
    1 & \qquad \text{if $\phi\ge\epsilon$},
  \end{array}
  \right.
\end{equation}
which has a continuous second derivative.
The momentum balance equation of the coupled fluid--structure system is
\begin{equation}
\label{eq:full_mmt}
\rho \left (\frac{\partial \vu}{\partial t}
 + ( \vu \cdot \nabla ) \vu \right) = - \nabla p +\nabla \cdot \vtau + \mathbf b
\end{equation}
with blended $\rho$, $\vtau$, and $\mathbf b$. The monolithic equation above satisfies the flow equation and elasticity equation, in the fluid and the solid phases, respectively.
Furthermore, the traction-matching condition is also automatically satisfied due to blending of stresses near the interface.
Eqs.~\eqref{eq:xi_adv}, \eqref{eq:incomp}, \& \eqref{eq:full_mmt}
form a single set of governing equations for the coupled FSI system.

In the solid bodies, we add an artificial viscosity $\mu_a$ to improve numerical stability.
If $\mu_a$ scales with grid size, in the limit of very fine spatial resolution, we recover the undamped solid equation.
If it is set to a grid size-independent constant, it is equivalent to simulating a Kelvin--Voigt viscoelastic solid.
We also define a dimensionless constant $\gamma_t$ that applies an additional multiplicative factor to the artificial viscosity in the blur zone (see SI for details).

%

%
\subsection*{Extrapolation of $\vxi(\vx,t )$}
Since $\vxi (\vx,t)$ is only defined inside the solid,
it needs to be extrapolated to several grid cells outside of the interface
for calculating derivatives in Eq.~\eqref{eq:xi_adv} and the deformation gradient tensor $\vF$ near the interface.
We describe our new extrapolation method for a single solid occupying domain $\Omega_{s}$, though it can be easily applied to any number of objects.
First we simplify the spatial order in which $\vxi(\vx, t)$ is extrapolated by making use of adjacency rules on a fixed grid.
Consider a Cartesian mesh with $M\times N\times O$ grid cells indexed by $i, j, k$,
we define orthogonal neighbors of a central cell as those cells that share a common face with the central cell (Fig.~\ref{fig:overview}(c)).
We label cells with $\phi < 0$ as the interior cells, or equivalently, as cells in the zeroth layer.
All the other cells are left as unmarked.
Then, for each interior cell, we label its unmarked orthogonal neighbors as the first layer cells, with index $l=1$.
We repeat this procedure to find subsequent layers $l=2,3,\ldots$, until we reach a physical boundary or the maximum number of layers, whichever occurs earlier.
The maximum number of layers is chosen so that the entire blue zone can be covered by the extrapolation procedure while conserving computational resources.

The extrapolation is then performed in ascending order of layers, but it can be computed independently for each cell within the same layer.
Suppose we aim to create an extrapolated reference map, $\vxi_e = (\xi_i, \xi_j, \xi_k)$ in layer $l_{i,j,k}$,
with cell index $(i,j,k)$. Subscript $e$ denotes an extrapolated cell.
Using WLS regression, we first build a local linear model of the reference map $\vxi(\vx, t)$ as a function of physical coordinate $\vx$.
To find eligible data points for the regression, we search within a $r_s=2$ box centered at cell $(i,j,k)$, where $r_s$ is the search radius measured in number of grid cells.
Consider a cell with index $(p,q,r)$ with $\vxi_d = (\xi_p, \xi_q, \xi_r)$, where subscript $d$ denotes a data cell.
It is eligible to be a data point in the regression only if it is marked as a layer in the previous procedure, and $l_{p,q,r} < l_{i,j,k}$.
In other words, reference map $\vxi_d$ must exist and be in layers lower than that of $\vxi_e$.
In the case of multiple solid objects, $\vxi_d$ must emanate from the same object whose extrapolation we seek.

The weights in the regression are important to ensure the quality of the extrapolated values,
especially when local deformations are large.
In the weighting scheme, we use an exponential decaying kernel centered at the extrapolated cell,
and near the interface we incorporate geometric information via $\phi_0$.
Details are provided in the SI.
The WLS problem is solved to obtain a set of coefficients $\vbeta$, and extrapolated reference map is calculated by $\vxi_e(\vx_e, t)=\vbeta \vx_e$.
If the linear system is degenerate, we increase the search radius $r_s$ by 1 and repeat the procedure from searching for eligible data points, but this is rare in practice.

In multi-body simulations, the extrapolation procedure is applied to each object independently.
We require that solid bodies do not co-exist at a grid cell.
However, the blur zone of an object is allowed to overlap with blur zones or interiors of other objects.
Thus, at a single grid cell, there can be several reference maps, extrapolated or not, each belonging to a distinct object.
Since extrapolated values are only needed in a small region near the interface, we design a custom data structure that is tailored to store these extrapolated values efficiently in memory.
Besides eliminating the need of reinitialization, the current extrapolation method has two additional advantages:
(1) layers can be defined given a definition of adjacency on the grid;
(2) the method is layer-wise and object-wise independent, thus easy to be parallelized.

\subsection*{Numerical procedures and implementation}
The RMT implementation in 3D (RMT3D) is developed in C\texttt{++} and parallelized via domain decomposition using the Message-Passing Interface (MPI) library.
The numerical schemes extend our previous work on 2D RMT implementation \cite{rycroft20}
and follow established discretizations for solving hyperbolic conservative laws \cite{almgren96, yu03}.
In summary, we extend from previous works the variable arrangement on the grid, finite-difference schemes to compute spatial derivatives,
and Godunov-type upwinding scheme for hyperbolic conservation laws \cite{colella90} to handle the advective parts of Eqs.~\eqref{eq:xi_adv} \& \eqref{eq:full_mmt}.
In addition, an approximate projection method \cite{almgren96, puckett97} and a marker-and-cell projection method
are used to enforce the incompressibility constraint (Eq.~\eqref{eq:incomp}) on the velocity solution at each timestep, and on an intermediate velocity field between two timesteps, respectively.
Large linear systems from the projection methods are solved using a custom geometric multigrid solver.
In many-body simulations, a collision stress-based contact model developed in the previous work \cite{rycroft20} is used.
Details of the numerical schemes and convergence tests are provided in the SI.

\section*{Results}
In this section, we consider immersed viscoelastic neo-Hookean solids (constant $\mu_a$) in various settings.
We nondimensionalize the governing equations using appropriate length, time, and mass scales in each test case.
We also use isotropic grid spacing $h$ in all simulations.

\subsection*{Settling sphere in a square cylinder}
            \begin{figure*}[ht]
            \centering
            \includegraphics[width=11.4cm,keepaspectratio]{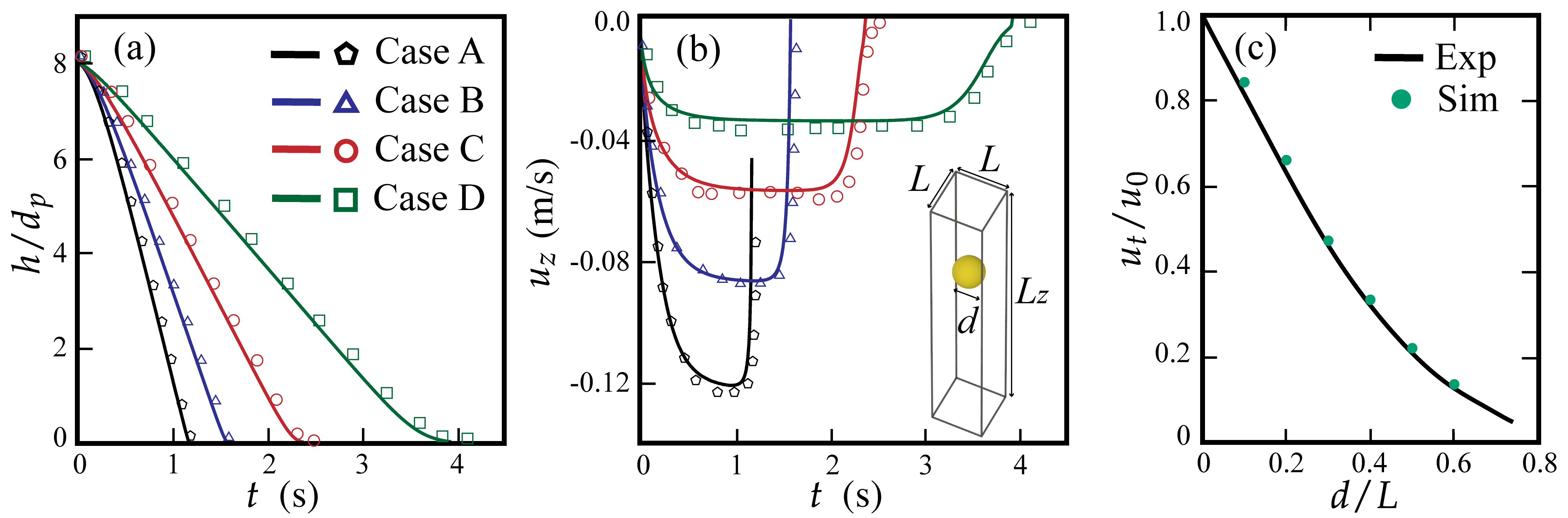}
            \caption{
		Comparisons of settling sphere simulations against experiments.
              The positions (a) and velocities (b) of a simulated sphere (solid lines) are compared with experimental results (symbols) by ten Cate \textit{et al.}~\cite{tenCate02}.
            Parameters are given in Table~\ref{table:params}.
            The inset in (b) shows an example simulation.
            (c) The wall correction factors in simulations are compared against experimental results by Miyamura \textit{et al.}~\cite{miyamura81}, using parameters $(L, L_z, \rho_s, \rho_f, G, \mu, \mu_a, \gamma_t, h, \epsilon)=(1, 6, 2, 1, 10, 0.04714, 0.04714, 0, 1/160, 1/320)$.
            Actual $\Rey_p=0.021,$ $0.132,$ $0.319,$ $0.537,$ $0.694,$ $0.743,$ respectively, for spheres with increasing diameters.}\label{fig:settling}
             \end{figure*}

The transport of rigid and deformable particles in fluid flow is central in many biological and physical systems.
While analytical solutions are available for simple cases such as a rigid sphere in unbound creeping flow, more complex FSIs in this setting elude analytical approaches.
In recent decades, the effects of confinement and multi-body interactions during settling have been a focus of much experimental and numerical work \cite{miyamura81, tenCate02, macmeccan09, van_der_sman10, ghosh15, alapati15}.
In this section, we simulate a solid sphere with a high shear modulus settling in a confined geometry.
We aim to demonstrate that by simply increasing the solid shear modulus, our method can easily capture the dynamics of a settling rigid sphere at various Reynolds number.

We compare simulated data with experimental measurements of positions, velocities, and wall correction factors on the terminal velocity of a rigid sphere settling in a square cylinder \cite{tenCate02, miyamura81}.
We devise a dimensionless parameter $\zeta \equiv \rho_f G d / (\rho_s \mu u_0)$ that compares the strength of the solid elastic stress against that of the viscous stress,
$u_0$ is the terminal velocity of a sphere in an unbound creeping flow, $G$ is the shear modulus, and $d$ is the sphere diameter.
We find that a moderate value of $G$, much smaller than the experimental values, suffices to satisfy $\zeta \gg 1$ and ensure minimum elastic deformation.
To accurately capture settling dynamics, it is critical to make sure that the artificial viscosity does not excessively add to the viscous drag at the interface.
There are two sources of additional drag in our method:
(1) viscous stress due to the use of $\mu_a$ and $\gamma_t$ in the blur zone
and
(2) solid shear stress blended into the fluid side of the blur zone.
To address (1) and still ensure stability, we restrict $\mu_a$ by $\mu_a/\mu \in [1,4]$.
In the meanwhile, we set $\gamma_t = \mu/\mu_a - 1$ and the blur zone width $\epsilon=0.5h$ so that the total viscosity on the fluid side is exactly $\mu$.
The solid elastic energy in all simulations in this section is $<8 \times 10^{-3}\%$ of the total energy,
confirming that the elastic deformation is negligible and so is the additional drag due to (2).

The simulation domain and results are shown in Fig.~\ref{fig:settling}.
In Table~\ref{table:params} we report the physical parameters used in experiments by ten Cate \textit{et al.}~\cite{tenCate02} and the corresponding dimensionless simulation parameters.
As shown in  Fig~\ref{fig:settling}(a,~b), positions and velocities in simulations agree well with all four experiments by ten Cate \textit{et al.}~\cite{tenCate02}.
To resolve the lubrication layer when the sphere approaches the bottom, we use an appropriate grid resolution ($h =1/160$) but no additional treatments.
In addition, we apply repulsive forces to the solid when it breaches a threshold distance to the bottom, to keep it from penetrating the wall (see SI for details).

The presence of walls drastically modifies the flow, reducing the terminal velocity from $u_0$ to $u_t$.
Define the ratio of the object size to the confinement size as $\eta = d/L$, where $L$ is the width of the square cylinder cross-section.
The reduction factor $u_t/u_0$ can be expressed as $u_t/u_0 = f(\eta)$, and it is measured experimentally by Miyamura \textit{et al.}~\cite{miyamura81}.
To match these results, we first nondimensionalize length, time, and mass using $l_0 = 0.1 \text{~m}, t_0 = 1 \text{~s}, m_0 = 10^{-3} \text{~kg}$, respectively.
The terminal velocity according to Stokes' law is $u_0 = {(\rho_s - \rho_f)gd^2}/{18\mu}$.
To emulate the flow conditions in these experiments, $\mu$ is chosen so that the particle Reynolds number $\Rey_p = \rho_s u_0 d / \mu$ is small.
The actual terminal velocity $u_t$ is averaged over time after it has been reached.
The apparatus used by Miyamura \textit{et al.}~has a height to width ratio of 100:1, which is hard to achieve in our simulations.
Instead we use a square cylinder with aspect ratio 6:1 and only consider data from the central $2/3$ of the domain.
Fig.~\ref{fig:settling}(c) shows that the wall correction factors in our simulations agree well with a reported curve fit of the experimental measurements \cite{miyamura81}.

\begin{figure*}[t]
            \centering
            \includegraphics[width=17.53cm,keepaspectratio]{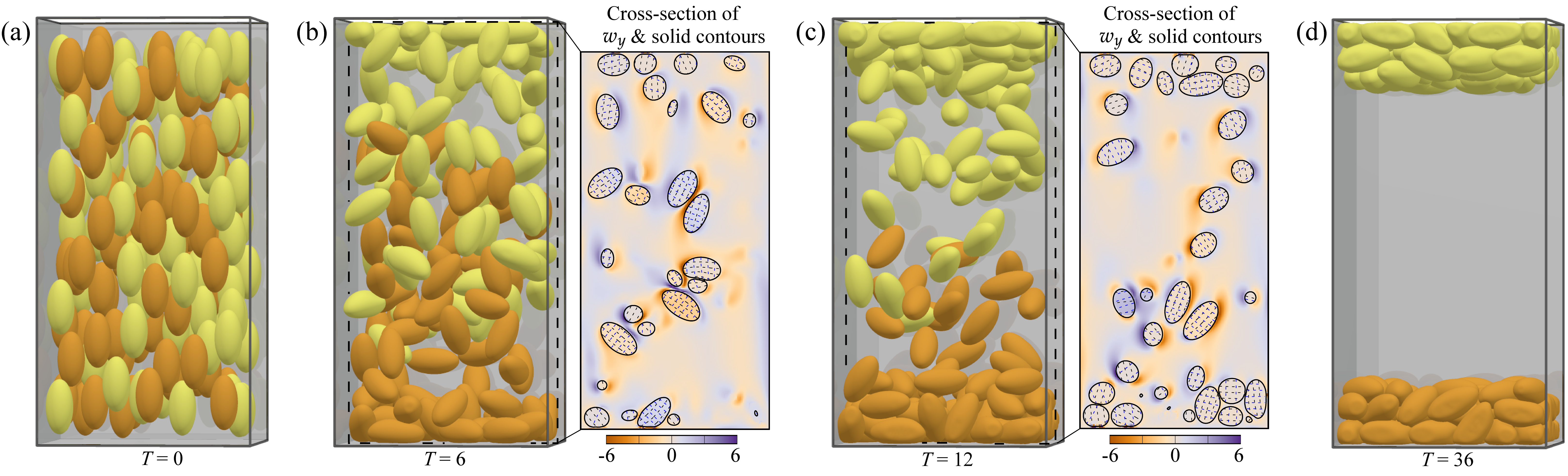}
            \caption{
             Various stages of 150 ellipsoids settling in a square cylinder (see SI for full movie).
             All ellipsoids have
             major axis $R=0.13$ initially aligned with $z$-direction, aspect ratio 2:1:1.
             Half of the ellipsoids are buoyant, $\rho_s/\rho_f = 0.8$ (yellow) and the others are denser than the fluid $\rho_s/\rho_f =1.25$ (orange).
             Other parameters are $(L_x, L_y, L_z, \rho_f, G, \mu, \mu_a, \gamma_t, h, \epsilon)=(1, 1, 2, 1, 10^{-1}, 10^{-3}, 10^{-3}, 0, 1/128, 1/128)$.
             No-slip boundary conditions are applied on all the walls.
             In (b) and (c) cross-sections at $y=0.5$ are shown, color corresponds to the $y$ component of vorticity.
             Fluid--solid interfaces are plotted with a thick black line, and contours of reference map components $\xi_x$ and $\xi_z$ are plotted with black and blue dashed lines, respectively.
             A contact model among solid bodies via collision stress~\cite{rycroft20} is used.
             The maximum particle Reynolds number is approximately 20.
             This simulation was run with 48 MPI processes on Intel ``Cascade Lake'' and took 19~hours to simulate to $T=36$.
             }\label{fig:sep_ells}
            \end{figure*}

Besides easily simulating a wide range of solid stiffness,
another benefit of the RMT is that buoyant or neutrally buoyant solids require no special treatment to address the added-mass effect, a numerical difficulty often suffered by partitioned FSI methods~\cite{causin05}.
This is an important advantage as many FSI problems of interest involve such density ratios in the solid and fluid phases, e.g.~problems in hemodynamics \cite{fai17a} and biomechanics \cite{kanso17}.
To demonstrate this, as well as the RMT's ability to simulate complex suspensions \cite{lindner14},
we show a simulation of the settling of 150 soft ellipsoids in Fig.~\ref{fig:sep_ells}.

             \begin{table*}
                \scriptsize
                \centering
                \caption{Physical parameters (columns 2--3) used in the experiments by ten Cate \textit{et al.}~\cite{tenCate02}
                and dimensionless simulation parameters (column 4--7).
                Experiment: sphere density $1120~\text{kg}/\text{m}^3$, diameter $15~\text{mm}$, and domain dimensions $10~\text{cm} \times 10~\text{cm} \times 16~\text{cm}$.
                Simulation: M, L, T are units of mass, length, and time, respectively.  Length and time are rescaled by $ l_0 = 0.1~\text{m}$ and $t_0 = 0.1~\text{s}$, respectively. Mass is rescaled in each case so that the fluid density is $1~\text{M/L}^3$.
                For all cases, $h = 1/160, \epsilon = 0.5h$.
                }\label{table:params}
                \begin{tabular} {| c | c | c || c | c | c | c |}
                    \hline
                   & $\rho_f$ & $\mu$  & $\rho_s$ & $\mu$  & $\mu_a/\mu$ & G   \\
                   & \tiny{($\text{kg/m}^3$)}  & \tiny{(kg/(ms))} & \tiny{($\text{M/L}^3$)}  & \tiny{($10^{-3}$M/(LT))} & -- & \tiny{($\text{M/LT}^2$)} \\
                    \hline \hline
                    A & 970 & 0.373 & 1.155 & 3.845 & 1 & 5.0 \\
                    \hline
                    B & 965 & 0.212 & 1.161 & 2.197 & 1& 2.5 \\
                    \hline
                    C & 962 & 0.113 & 1.164 & 1.175 & 2 & 2.0 \\
                    \hline
                    D & 960 & 0.058 & 1.167 & 0.6042 & 4 & 2.0 \\
                    \hline
                \end{tabular}
            \end{table*}


\subsection*{Lid-driven cubic cavity with a sphere}
\begin{figure*}[t]
  \centering
  \includegraphics[width=17.8cm,keepaspectratio]{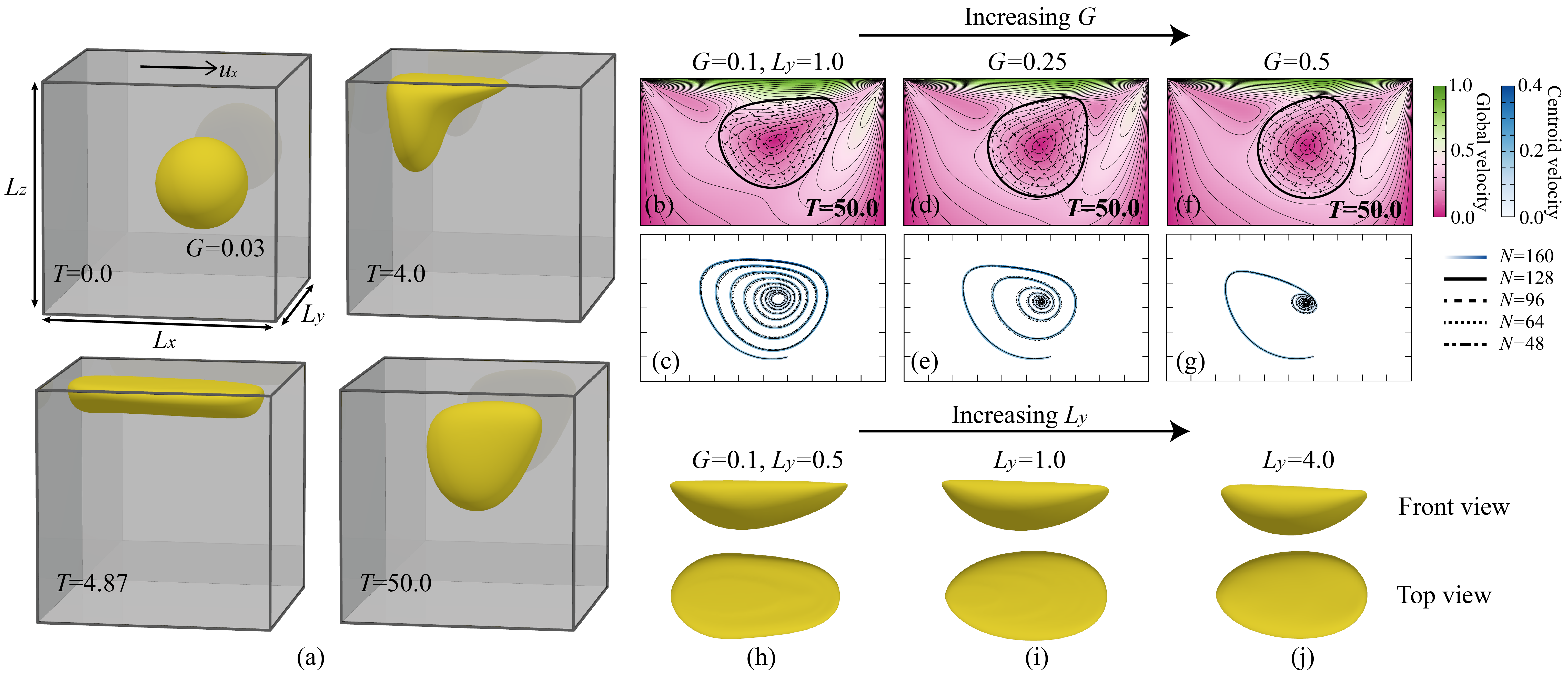}
   \caption{
   Simulations of a sphere in the lid-driven cavity.
  For all cases, parameters $(L_x, L_z, u_x, \rho_s, \rho_f,  \mu, \mu_a, \epsilon)=(1, 1, 1, 1, 1, 10^{-2}, 10^{-2}, 1.5h), \gamma_t \in [0,4], h \in [1/160, 1/48]$.
  The sphere has radius $0.2$ and is initially centered at $(0.6, 0.5L_y, 0.5)$.
  (a) 3D snapshots of a simulation at various times, $(L_y, G) = (1, 0.03)$.
  (b) Contours of the reference map and the interface are plotted against a background of velocity magnitude (heatmap and contours).
  A cross-section at $y=0.5$ and $T=50$ is shown, $(L_y, G) = (1, 0.1)$, $h = 1/N$.
  (c) The trajectory of the sphere centroid from $T=0$ to $T=50$ of the sphere in (b).
  The color corresponds to the velocity magnitude of the centroid.
  (d), (f) The same as (b) but for solid shear moduli $G=0.25$ and $0.5$, respectively.
  (e), (g) The same as (c) but for solid shear moduli $G=0.25$ and $0.5$, respectively.
  (h), (i), (j) 3D shape of a deformed $G=0.1$ sphere at the closest approach to the top in cavity with $L_y=0.5, 1.0, 4.0$, respectively.
  See SI for additional parameters, movies, and trajectory data.
  }
  \label{fig:lid_driven_panel}
 \end{figure*}

In computational fluid dynamics the lid-driven cavity has long been an important benchmark problem \cite{albensoeder05, gelfgat19}.
Despite its simplicity, it exhibits rich flow dynamics due to varying cavity geometries, boundary conditions, and Reynolds numbers.
In stark contrast to the extensive studies on the fluid problem in both 2D and 3D, results on lid-driven cavities with deformable boundaries and immersed solids are much fewer \cite{dunne06, zhao08, zhang13, villone19}.
In this section, we turn our attention to a neutrally buoyant deformable sphere in a lid-driven cavity,
which to our best knowledge is the first 3D result of its kind.

Shown in Fig.~\ref{fig:lid_driven_panel}(a), the cavity has size $L_x \times L_y \times L_z$ and two span aspect ratios, $ L_z/L_x$ and $L_y/L_x$.
The lid moves with velocity $\vu_\text{top} = (u_x,0,0)$ and no-slip boundary conditions are applied on all the other walls.
We rescale length and velocity by $L_x$ and $u_x$, respectively.
As a validation, we simulate lid-driven cavity flows without a solid, configured with various span aspect ratios and Reynolds numbers.
Our results agree well with high accuracy benchmarks \cite{albensoeder05} (see SI for details).

A circular particle in a square lid-driven cavity in 2D has been investigated by Zhao \textit{et al.}~\cite{zhao08} and widely used as a validation case in later works \cite{sugiyama11, esmailzadeh14, farahbakhsh16, casquero18}.
We simulate a sphere in a cubic cavity but choose parameters similar to those in the 2D test case to highlight qualitative differences in 3D (Fig.~\ref{fig:lid_driven_panel}(b,~c)).
The middle cross-section of the deformed sphere (Fig.~\ref{fig:lid_driven_panel}(b)) is qualitatively similar to the shape of the deformed circular particle at long time \cite{casquero18}.
The distinctions between the two cases are more apparent in their centroid trajectories.
Compared with the 2D case \cite{sugiyama11}, although the centroid of a sphere in a cubic cavity (Fig.~\ref{fig:lid_driven_panel}(c)) also converges to a stationary point, each spiral of its trajectory is much closer to the neighboring ones.
There are several reasons for this, most notably the topological difference between an infinite cylinder and a sphere.
Another reason is the reduced circulation due to lateral walls in the third dimension \cite{gelfgat19},
which allows the sphere to interact with the moving lid for a longer time before being carried back to the center of the cavity by the flow.

We also simulate spheres with varying shear moduli.
Snapshots of a simulation with $G=0.03$ are shown in Fig.~\ref{fig:lid_driven_panel}(a).
To our knowledge, this is the lowest shear modulus reported in the literature for cavity flow with deformable solids.
In addition, as Fig.~\ref{fig:lid_driven_panel}(e,~g) show,
as the shear modulus increases, the sphere exhibits distinct centroid trajectories.
Fig.~\ref{fig:lid_driven_panel}(d,~f) offer some intuition for the qualitative changes.
As a stiffer sphere moves toward and along the top lid,
it deforms less and is able to separate earlier from driving flow.
Consequently, a stiffer sphere is carried by the flow from the top right more toward the center than toward the bottom.

We also want to show the effect of the lateral walls on the solid deformation.
We simulate a sphere with $G=0.1$ initially at $(0.6, 0.5L_y, 0.5)$ in cavities with $L_y/L_x=0.5, 4$.
Fig. \ref{fig:lid_driven_panel}(h--j) show two views of the sphere at the closest approach to the top lid in each cavity.
As the walls become farther apart, the sphere becomes less stretched lengthwise and less compressed vertically,
again suggesting that a reduced circulation increases the strength of the interaction between the sphere and the moving boundary.
We note that here we do not impose any repulsive forces on the sphere near the walls to avoid interfering with its dynamics,
thus properly resolving the lubrication layer between the sphere and the boundaries is critical in keeping it from penetrating the walls.
In simulations with a coarse resolution and a low $G$, a nonzero $\gamma_t$ is needed to dampen the motion of the interface near the boundaries.

In this section, we demonstrate that our numerical schemes, including the new extrapolation method, perform well under the stress of simulating extreme deformations near boundary singularities in a dynamic flow.

\subsection*{Swimming}
\begin{figure*}[ht]
    \centering
    \includegraphics[width=11.4cm,keepaspectratio]{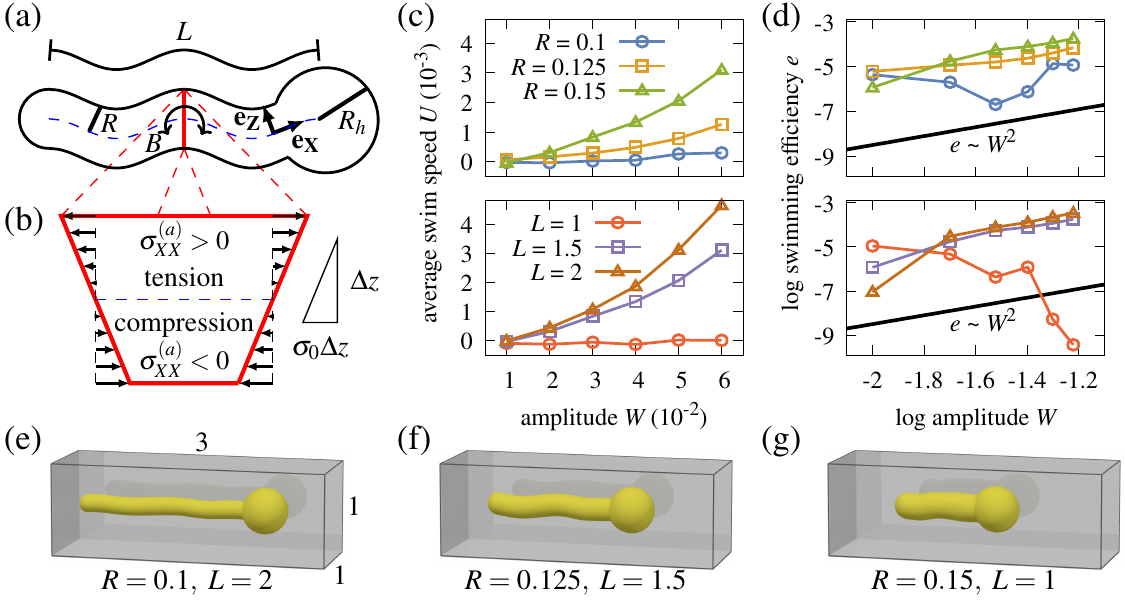}
    \caption{
    Swimming of flagellar objects driven by active bending moments.
    (a) Schematic of a swimmer. Swimmers have a spherical head of radius $R_h$ and a cylindrical body of length $L$ and a body radius $R < R_h$.
    The body is actuated by traveling wave of bending moments $B$ applied to vertical cross-sections (marked in red).
    (b) Expanded view showing the active in-plane stresses $\sigma^{(a)}_{XX}$ directed along the body's long axis $\Xhat$.
    The stress magnitude varies linearly with the height above the horizontal midplane (a dashed blue line).
    The swim speed $U$ (c) and efficiency $e$ (d) as a function of $W$, the bending wave amplitude, are presented for a variety of body shapes.
    Simulations in the top row vary $R$ with $L=1.5$, and those in the bottom row vary $L$ with $R=0.15$.
    (e--g) Body shapes, mid-stroke, are shown for several $R,L$ combinations. See SI for simulation movies.
    Steady-state measurements are taken at time $T$.
    $T=20$ for $R=0.15$, and $T=10$ otherwise.
    For all cases, $(L_x, L_y, L_z, R_h, \rho,G, \mu,\mu_a,\gamma_t, h, \epsilon) = (3,1,1, 0.25,1,1, 10^{-2},10^{-2},0, 1/96, 1/192)$.}
    \label{fig:swimming}
\end{figure*}

We now apply the RMT to model swimming where the active deformations in the solid phase drive the motions in the system.
Swimming has been an FSI problem of interest for decades \cite{taylor51,purcell77,lighthill1976}. To address the difficulty of resolving the full FSI, especially motions of the phase boundary, modelers often apply simplifications such as asymptotic analysis and scaling arguments.
A related, widely applied numerical approach in the low Reynolds number regime is to abstract the swimmer into a one-dimensional collection of regularized singularities \cite{lim2019, olson2020, shelley2000, medovikov2005, leiderman2016,fauci2017}.
As the swimmer undergoes prescribed deformation or active forcing, the resulting flow is a good approximation to the swimmer's far field.
However, this method is not well suited for contexts where the near field is of prime importance, such as dense suspensions of swimmers and swimming in tight confinement.
These cases often require specializations \cite{nguyen2019,gaffney2019} or immersed boundary methods \cite{kanso2018,arash2020}.
Here, we show that the RMT naturally resolves the near field around finite-size swimmers in confined geometries.
Despite being an Eulerian method, the RMT provides easy access to the reference map, allowing for straightforward definition of the active stresses in the swimmer's body frame.


We begin with a description of our model swimmer, a cylindrical flagellum of length $L$ and radius $R$ with a spherical head of radius $R_h > R$ as shown in Fig.~\ref{fig:swimming}(a), and the active stress driving its cyclic deformation.
We decompose the solid stress tensor into passive and active parts $\vsigma_s = \vsigma^{(\text{p})} + \vsigma^{(\text{a})}$,
where $\vsigma^{(\text{p})} = \vsigma^{(\text{p})}(\vF)$ is the elastic stress tensor from previous sections.
We seek to define a time-dependent active stress field $\vsigma^{(\text{a})} = \vsigma^{(\text{a})}(t,\vxi,\vF)$ as a function of body position and local deformation,
which will induce a planar bending wave traveling along the cylinder flagellum.
First, we specify how body orientation is determined in the deformed frame.
We assume without loss of generality that the reference frame coordinate system is centered on the flagellum and aligned with the swimmer.
Denoting the reference frame's orthonormal basis as $\{\Ihat\}$, $I \in {X, Y, Z}$,
we orient $\{\Ihat\}$ so $\Xhat$ points along the body toward the head and $\Zhat$ vertically up.
The directions of these vectors in the deformed space, $\{\ihat\}$,
where $\smash{\ihat = \vF \cdot \Ihat / \|\vF \cdot \Ihat\|}$,
are also body-aligned as shown in Fig.~\ref{fig:swimming}(a).

Now, we define the active stress in terms of the reference map coordinates $\smash{X(\vx,t) = \Xhat \cdot \vxi(\vx,t)}$ and $\smash{Z(\vx,t) = \Zhat \cdot \vxi(\vx,t)}$, which denote the reference distance along the cylinder and above the midplane, respectively.
Bending moments are induced by axial stresses $\smash{\sigma_{XX}^{(a)} = \xhat \cdot \vsigma^{(a)} \cdot \xhat}$ of opposing sign about the midplane $Z=0$, as shown in Fig.~\ref{fig:swimming}(b), so we set $\sigma_{XX}^{(a)} = \sigma_0 Z$ for simplicity.
Letting the full tensor be traceless, we set
\begin{equation}
    \vsigma^{(\text{a})} = \frac{3Z\sigma_0}{2} \left(\xhat \otimes \xhat - \frac13 \vI\right),
\end{equation}
where $\sigma_0 \propto \cos(kX - \omega t)$ for some wavenumber $k$ and frequency $\omega$.
Introducing the amplitude parameter $W = B/3GI k^2$, where $I$ is the cylinder's cross-sectional area moment of inertia and $B$ a bending moment magnitude, we set
\begin{equation}
    \sigma_0 = 3 W G k^2 \cos(k X - \omega t),
\end{equation}
to systematically set bending moments across a range of $R$.
We emphasize that $W$ is not a prescribed or measured vertical displacement, but a parameter derived from scaling arguments describing linear rod bending (see SI for details).

We nondimensionalize variables and parameters using length, time, and stress scales $l = 2\pi/k$, $\tau = 2\pi/\omega$, $\Sigma = G$.
We simulate the dimensionless, uniform density system at constant $\rho, \mu$ and $R_h$ with varying $R$, $W$, and $L$.
For each combination of parameters, we calculate the average swim speed $U$, shown in Fig.~\ref{fig:swimming}(c), and active power $\smash{P= \big<-\int_{\Omega_s} \vsigma^{(a)}:\nabla \vu dV\big>}$, where $\left<*\right>$ denotes time averaging over many cycles of oscillation. 
We also calculate an approximate Lighthill efficiency $e = CU^2/P$ \cite{lighthill1975}, shown in Fig.~\ref{fig:swimming}(d), where $C$ is a  drag coefficient used to estimate the force required to tow the swimmer at velocity $U$.
Mid-stroke body shapes for selected parameters are shown in Fig.~\ref{fig:swimming}(e--g). See SI for details and movies.

Broadly, swimming with the prescribed active stress is faster and more efficient at larger body sizes.
This may be in part due to the changing Reynolds number $\Rey_s$ corresponding to the time-averaged object motion, since it grows as $\Rey_s \sim W^2$ over the range of simulated $W$ values; in contrast, the Reynolds number $\Rey_o$ describing the swimming gait varies more slowly as $\Rey_o\sim W$. Both $\Rey_s,\Rey_o \in [10^{-4},0.1]$ for the results presented in Fig.~\ref{fig:swimming} (see SI for details.)
At large amplitudes, the efficiency scales as $e \sim W^2$, consistent with \smash{$U, P \sim W^2$}.
Notably, there is effectively no motion at $L=1$, suggesting a minimum length is required for efficient locomotion.

\section*{Conclusion and future work}
The reference map technique is an efficient and flexible numerical method for FSI problems that involve many bodies, complex solid geometries, large deformations, and actuation.
We have demonstrated the accuracy of its first 3D implementation, RMT3D, by convergence tests and comparisons against experimental data of settling spheres.
We have also presented its applications to simulate settling of a large number of ellipsoids with varying density, a soft sphere in several lid-driven cavity flows, and swimmers with different body geometries, actuated by active stress.

A major future direction for the RMT is to address the issue of solid self-contact,
which is a common occurrence in geometrically large deforming bodies, such as long slender structures and active swimmers.
Other future work includes developing a more accurate contact model with physical boundaries to capture rebound behavior,
as well as applying adaptive mesh refinement techniques to increase computational efficiency in many-body and multi-scale systems.

\section*{Acknowledgement}
Y.L.L. acknowledges support from the Department of Energy Computational Sciences Graduate Fellowship program.
N.J.D. acknowledges support from the Department of Defense NDSEG Fellowship. Y.L.L. and N.J.D. are also supported by the Harvard NSF-Simons Center Quantitative Biology Initiative student fellowship.
C.H.R. was partially supported by the Applied Mathematics Program of the
U.S. DOE Office of Science Advanced Scientific Computing Research under
contract number DE-AC02-05CH11231.\\

\noindent
\textbf{Author contributions}\\
Y.~L.~L., N.~J.~D. and C.~H.~R. designed and performed research, wrote simulation codes, analyzed data, and wrote the paper.\\

\noindent
\textbf{Competing interest}\\
The authors declare no conflict of interest.\\

\noindent
\textbf{Code availability}\\
The simulation codes, RMT3D, are  available on GitHub at
\url{https://github.com/ylunalin/rmt3D}
\\
The custom linear solver, Parallelized Geometric Multigrid (PGMG), which is required by RMT3D, is available on GitHub at
\url{https://github.com/chr1shr/pgmg}

\bibliographystyle{unsrt}
\bibliography{references}

\begin{thebibliography}{10}

\bibitem{luhar11}
Mitul Luhar and Heidi Nepf.
\newblock Flow-induced reconfiguration of buoyant and flexible aquatic
  vegetation.
\newblock {\em Limnology and Oceanography}, 56:2003--2017, 11 2011.

\bibitem{bukowicki19}
Marek Bukowicki and Maria~L. Ekiel-Je\.{z}ewska.
\newblock Sedimenting pairs of elastic microfilaments.
\newblock {\em Soft Matter}, 15(46):9405--9417, 2019.

\bibitem{peskin72}
Charles~S Peskin.
\newblock Flow patterns around heart valves: A numerical method.
\newblock {\em Journal of Computational Physics}, 10(2):252--271, 1972.

\bibitem{griffith09}
Boyce~E. Griffith, Xiaoyu Luo, David~M. McQueen, and Charles~S. Peskin.
\newblock Simulating the fluid dynamics of natural and prosthetic heart valves
  using the immersed boundary method.
\newblock {\em International Journal of Applied Mechanics}, 01(01):137--177,
  2009.

\bibitem{lucas14}
Kelsey~N. Lucas, Nathan Johnson, Wesley~T. Beaulieu, Eric Cathcart, Gregory
  Tirrell, Sean~P. Colin, Brad~J. Gemmell, John~O. Dabiri, and John~H.
  Costello.
\newblock Bending rules for animal propulsion.
\newblock {\em Nature Communications}, 5(1), December 2014.

\bibitem{lim2019}
Yunyoung Park, Yongsam Kim, and Sookkyung Lim.
\newblock Locomotion of a single-flagellated bacterium.
\newblock {\em Journal of Fluid Mechanics}, 859:586--612, 2019.

\bibitem{kanso17}
Janna~C. Nawroth, Hanliang Guo, Eric Koch, Elizabeth A.~C. Heath-Heckman,
  John~C. Hermanson, Edward~G. Ruby, John~O. Dabiri, Eva Kanso, and Margaret
  McFall-Ngai.
\newblock Motile cilia create fluid-mechanical microhabitats for the active
  recruitment of the host microbiome.
\newblock {\em Proceedings of the National Academy of Sciences},
  114(36):9510--9516, 2017.

\bibitem{kanso2018}
Hanliang Guo, Lisa Fauci, Michael Shelley, and Eva Kanso.
\newblock Bistability in the synchronization of actuated microfilaments.
\newblock {\em Journal of Fluid Mechanics}, 836:304--323, 2018.

\bibitem{peskin02}
Charles~S. Peskin.
\newblock The immersed boundary method.
\newblock {\em Acta Numerica}, 11:479--517, 2002.

\bibitem{fai18}
Thomas~G. Fai and Chris~H. Rycroft.
\newblock Lubricated immersed boundary method in two dimensions.
\newblock {\em J. Comput. Phys.}, 356:319--339, 2018.

\bibitem{griffith20}
Boyce~E. Griffith and Neelesh~A. Patankar.
\newblock Immersed methods for fluid--structure interaction.
\newblock {\em Annual Review of Fluid Mechanics}, 52(1):421--448, 2020.

\bibitem{hirt74}
Cyril~W. Hirt, Anthony~A. Amsden, and J.~L. Cook.
\newblock An arbitrary {L}agrangian {E}ulerian computing method for all flow
  speeds.
\newblock {\em J. Comput. Phys.}, 14:227--253, 1974.

\bibitem{rugonyi01}
Sandra Rugonyi and Klaus-J\"urgen Bathe.
\newblock On finite element analysis of fluid flows fully coupled with
  structural interactions.
\newblock {\em Comput. Model. Eng. Sci.}, 2:195--212, 2001.

\bibitem{truesdell55}
Clifford Truesdell.
\newblock Hypo-elasticity.
\newblock {\em Indiana Univ. Math. J.}, 4:83--133, 1955.

\bibitem{udaykumar03}
H~S Udaykumar, L~Tran, D~M Belk, and K~J Vanden.
\newblock An {Eulerian} method for computation of multimaterial impact with
  {ENO} shock-capturing and sharp interfaces.
\newblock {\em Journal of Computational Physics}, 186(1):136--177, 2003.

\bibitem{rycroft12}
Chris~H. Rycroft and Fr\'ed\'eric Gibou.
\newblock Simulations of a stretching bar using a plasticity model from the
  shear transformation zone theory.
\newblock {\em Journal of Computational Physics}, 231(5):2155--2179, 2012.

\bibitem{maitre09}
Emmanuel Maitre, Thomas Milcent, Georges-Henri Cottet, Annie Raoult, and Yves
  Usson.
\newblock Applications of level set methods in computational biophysics.
\newblock {\em Mathematical and Computer Modelling}, 49(11--12):2161--2169,
  2009.
\newblock Trends in Application of Mathematics to Medicine.

\bibitem{liu01}
Chun Liu and Noel~J. Walkington.
\newblock An {E}ulerian description of fluids containing visco-elastic
  particles.
\newblock {\em Archive for Rational Mechanics and Analysis}, 159(3):229--252,
  2001.

\bibitem{sugiyama11}
Kazuyasu Sugiyama, Satoshi Ii, Shintaro Takeuchi, Shu Takagi, and Yoichiro
  Matsumoto.
\newblock A full {E}ulerian finite difference approach for solving
  fluid--structure coupling problems.
\newblock {\em J. Comput. Phys.}, 230(3):596--627, 2011.

\bibitem{dunne06}
Thomas Dunne.
\newblock An {E}ulerian approach to fluid--structure interaction and
  goal-oriented mesh adaptation.
\newblock {\em Int. J. Numer. Methods Fluids}, 51(9-10):1017--1039, 2006.

\bibitem{richter13}
Thomas Richter.
\newblock A fully {E}ulerian formulation for fluid--structure interaction
  problems.
\newblock {\em J. Comput. Phys.}, 233:227--240, 2013.

\bibitem{wick13}
Thomas Wick.
\newblock Fully {E}ulerian fluid--structure interaction for time-dependent
  problems.
\newblock {\em Comput. Method. Appl. M.}, 255:14--26, 2013.

\bibitem{kamrin_thesis}
Ken Kamrin.
\newblock {\em Stochastic and Deterministic Models for Dense Granular Flow}.
\newblock PhD thesis, MIT, 2008.

\bibitem{kamrin12}
Ken Kamrin, Chris~H. Rycroft, and Jean-Christophe Nave.
\newblock Reference map technique for finite-strain elasticity and fluid--solid
  interaction.
\newblock {\em J. Mech. Phys. Solids}, 60(11):1952--1969, 2012.

\bibitem{valkov15}
Boris Valkov, Chris~H. Rycroft, and Ken Kamrin.
\newblock Eulerian method for multiphase interactions of soft solid bodies in
  fluids.
\newblock {\em J. Appl. Mech.}, 82(4):041011, 04 2015.

\bibitem{rycroft20}
Chris~H Rycroft, Chen-Hung Wu, Yue Yu, and Ken Kamrin.
\newblock Reference map technique for incompressible fluid--structure
  interaction.
\newblock {\em Journal of Fluid Mechanics}, 898:A9, 2020.

\bibitem{osher88}
Stanley Osher and James~A. Sethian.
\newblock Fronts propagating with curvature-dependent speed: Algorithms based
  on {H}amilton--{J}acobi formulations.
\newblock {\em J. Comput. Phys.}, 79(1):12--49, 1988.

\bibitem{sethian99}
James~A. Sethian.
\newblock {\em Level Set Methods and Fast Marching Methods: Evolving interfaces
  in computational geometry, fluid mechanics, computer vision and materials
  science}.
\newblock Cambridge University Press, 1999.

\bibitem{macminn15}
Christopher~W. MacMinn, Eric~R. Dufresne, and John~S. Wettlaufer.
\newblock Fluid-driven deformation of a soft granular material.
\newblock {\em Phys. Rev. X}, 5:011020, Feb 2015.

\bibitem{tytell2010}
Eric~D. Tytell, Chia-Yu Hsu, Thelma~L. Williams, Avis~H. Cohen, and Lisa~J.
  Fauci.
\newblock Interactions between internal forces, body stiffness, and fluid
  environment in a neuromechanical model of lamprey swimming.
\newblock {\em Proc. Natl. Acad. Sci.}, 107(46):19832--19837, 2010.

\bibitem{thomases2014}
Becca Thomases and Robert~D. Guy.
\newblock Mechanisms of elastic enhancement and hindrance for finite-length
  undulatory swimmers in viscoelastic fluids.
\newblock {\em Phys. Rev. Lett.}, 113:098102, Aug 2014.

\bibitem{olson2020}
Cole Jeznach and Sarah~D Olson.
\newblock Dynamics of swimmers in fluids with resistance.
\newblock {\em Fluids (Basel)}, 5(1):14, 2020.

\bibitem{gurtin10}
Morton~E. Gurtin, Eliot Fried, and Lallit Anand.
\newblock {\em The Mechanics and Thermodynamics of Continua}.
\newblock Cambridge University Press, 2010.

\bibitem{sussman94}
Mark Sussman, Peter Smereka, and Stanley Osher.
\newblock A level set approach for computing solutions to incompressible
  two-phase flow.
\newblock {\em J. Comput. Phys.}, 114(1):146--159, 1994.

\bibitem{sussman99}
Mark Sussman, Ann~S Almgren, John~B Bell, Phillip Colella, Louis~H Howell, and
  Michael~L Welcome.
\newblock An adaptive level set approach for incompressible two-phase flows.
\newblock {\em Journal of Computational Physics}, 148(1):81--124, 1999.

\bibitem{yu03}
Jiun-Der Yu, Shinri Sakai, and James~A. Sethian.
\newblock A coupled level set projection method applied to ink jet simulation.
\newblock {\em {Interfaces and Free Boundaries}}, 5(4):459--482, 2003.

\bibitem{almgren96}
Ann~S. Almgren, John~B. Bell, and William~G. Szymczak.
\newblock A numerical method for the incompressible {Navier--Stokes} equations
  based on an approximate projection.
\newblock {\em SIAM J. Sci. Comput.}, 17(2):358--369, 1996.

\bibitem{colella90}
Phillip Colella.
\newblock Multidimensional upwind methods for hyperbolic conservation laws.
\newblock {\em Journal of Computational Physics}, 87(1):171--200, 1990.

\bibitem{puckett97}
Elbridge~Gerry Puckett, Ann~S. Almgren, John~B. Bell, Daniel~L. Marcus, and
  William~J. Rider.
\newblock A high-order projection method for tracking fluid interfaces in
  variable density incompressible flows.
\newblock {\em J. Comput. Phys.}, 130(2):269--282, 1997.

\bibitem{tenCate02}
A.~ten Cate, C.~H. Nieuwstad, J.~J. Derksen, and H.~E.~A. Van~den Akker.
\newblock Particle imaging velocimetry experiments and lattice-{Boltzmann}
  simulations on a single sphere settling under gravity.
\newblock {\em Physics of Fluids}, 14(11):4012--4025, nov 2002.

\bibitem{miyamura81}
A.~Miyamura, S.~Iwasaki, and T.~Ishii.
\newblock Experimental wall correction factors of single solid spheres in
  triangular and square cylinders, and parallel plates.
\newblock {\em International Journal of Multiphase Flow}, 7(1):41--46, feb
  1981.

\bibitem{macmeccan09}
Robert~M. MacMeccan, J.~R. Clausen, G.~P. Neitzel, and C.~K. Aidun.
\newblock Simulating deformable particle suspensions using a coupled
  lattice-{Boltzmann} and finite-element method.
\newblock {\em Journal of Fluid Mechanics}, 618:13--39, January 2009.

\bibitem{van_der_sman10}
R.G.M. van~der Sman.
\newblock Drag force on spheres confined on the center line of rectangular
  microchannels.
\newblock {\em Journal of Colloid and Interface Science}, 351(1):43--49,
  November 2010.

\bibitem{ghosh15}
Sudeshna Ghosh and John~M. Stockie.
\newblock Numerical simulations of particle sedimentation using the immersed
  boundary method.
\newblock {\em Communications in Computational Physics}, 18(2):380--416, August
  2015.

\bibitem{alapati15}
Suresh Alapati, Woo~Seong Che, and Yong~Kweon Suh.
\newblock Simulation of {Sedimentation} of a {Sphere} in a {Viscous} {Fluid}
  {Using} the {Lattice} {Boltzmann} {Method} {Combined} with the {Smoothed}
  {Profile} {Method}.
\newblock {\em Advances in Mechanical Engineering}, 7(2):794198, February 2015.

\bibitem{causin05}
P.~Causin, J.F. Gerbeau, and F.~Nobile.
\newblock Added-mass effect in the design of partitioned algorithms for
  fluid--structure problems.
\newblock {\em Computer Methods in Applied Mechanics and Engineering},
  194(42-44):4506--4527, October 2005.

\bibitem{fai17a}
Thomas~G. Fai, Alejandra Leo-Macias, David~L. Stokes, and Charles~S. Peskin.
\newblock Image-based model of the spectrin cytoskeleton for red blood cell
  simulation.
\newblock {\em PLOS Computational Biology}, 13(10):1--25, 10 2017.

\bibitem{lindner14}
Anke Lindner.
\newblock Flow of complex suspensions.
\newblock {\em Physics of Fluids}, 26(10):101307, 2014.

\bibitem{albensoeder05}
S.~Albensoeder and H.C. Kuhlmann.
\newblock Accurate three-dimensional lid-driven cavity flow.
\newblock {\em Journal of Computational Physics}, 206(2):536--558, July 2005.

\bibitem{gelfgat19}
Hendrik~C. Kuhlmann and Francesco Roman\`{o}.
\newblock The {Lid}-{Driven} {Cavity}.
\newblock In Alexander Gelfgat, editor, {\em Computational {Modelling} of
  {Bifurcations} and {Instabilities} in {Fluid} {Dynamics}}, volume~50, pages
  233--309. Springer International Publishing, Cham, 2019.
\newblock Series Title: Computational Methods in Applied Sciences.

\bibitem{zhao08}
Hong Zhao, Jonathan~B. Freund, and Robert~D. Moser.
\newblock A fixed-mesh method for incompressible flow--structure systems with
  finite solid deformations.
\newblock {\em Journal of Computational Physics}, 227(6):3114--3140, March
  2008.

\bibitem{zhang13}
Zhi-Qian Zhang, G.~R. Liu, and Boo~Cheong Khoo.
\newblock A three dimensional immersed smoothed finite element method ({3D}
  {IS}-{FEM}) for fluid--structure interaction problems.
\newblock {\em Computational Mechanics}, 51(2):129--150, February 2013.

\bibitem{villone19}
Massimiliano~M. Villone and Pier~Luca Maffettone.
\newblock Dynamics, rheology, and applications of elastic deformable particle
  suspensions: a review.
\newblock {\em Rheologica Acta}, 58(3-4):109--130, April 2019.

\bibitem{esmailzadeh14}
H.~Esmailzadeh and M.~Passandideh-Fard.
\newblock Numerical and {Experimental} {Analysis} of the {Fluid}-{Structure}
  {Interaction} in {Presence} of a {Hyperelastic} {Body}.
\newblock {\em Journal of Fluids Engineering}, 136(11):111107, November 2014.

\bibitem{farahbakhsh16}
Iman Farahbakhsh, Hassan Ghassemi, and Fereidoun Sabetghadam.
\newblock A vorticity based approach to handle the fluid--structure interaction
  problems.
\newblock {\em Fluid Dynamics Research}, 48(1):015509, February 2016.

\bibitem{casquero18}
Hugo Casquero, Yongjie~Jessica Zhang, Carles Bona-Casas, Lisandro Dalcin, and
  Hector Gomez.
\newblock Non-body-fitted fluid--structure interaction: Divergence-conforming
  {B}-splines, fully-implicit dynamics, and variational formulation.
\newblock {\em Journal of Computational Physics}, 374:625--653, December 2018.

\bibitem{taylor51}
Geoffrey Taylor.
\newblock Analysis of the swimming of microscopic organisms.
\newblock {\em Proceedings of the Royal Society of London. Series A,
  Mathematical and physical sciences}, 209(1099):447--461, 1951.

\bibitem{purcell77}
E.~M Purcell.
\newblock Life at low {Reynolds} number.
\newblock {\em American Journal of Physics}, 45(1):3--11, 1977.

\bibitem{lighthill1976}
James Lighthill.
\newblock Flagellar hydrodynamics.
\newblock {\em SIAM Review}, 18(2):161--230, 1976.

\bibitem{shelley2000}
Michael~J Shelley and Tetsuji Ueda.
\newblock The {Stokesian} hydrodynamics of flexing, stretching filaments.
\newblock {\em Physica D}, 146(1):221--245, 2000.

\bibitem{medovikov2005}
Ricardo Cortez, Lisa Fauci, and Alexei Medovikov.
\newblock The method of regularized {Stokeslets} in three dimensions: Analysis,
  validation, and application to helical swimming.
\newblock {\em Physics of Fluids}, 17(3):031504--031504--14, 2005.

\bibitem{leiderman2016}
Karin Leiderman and Sarah~D Olson.
\newblock Swimming in a two-dimensional {Brinkman} fluid: Computational
  modeling and regularized solutions.
\newblock {\em Physics of Fluids}, 28(2):21902, 2016.

\bibitem{fauci2017}
Qiang Yang and Lisa Fauci.
\newblock Dynamics of a macroscopic elastic fibre in a polymeric cellular flow.
\newblock {\em Journal of Fluid Mechanics}, 817:388--405, 2017.

\bibitem{nguyen2019}
Hoang-Ngan Nguyen, Hoang-Ngan Nguyen, Sarah~D Olson, Sarah~D Olson, Karin
  Leiderman, and Karin Leiderman.
\newblock Computation of a regularized {Brinkmanlet} near a plane wall.
\newblock {\em Journal of Engineering Mathematics}, 114(1):19--41, 2019.

\bibitem{gaffney2019}
B.~J. Walker, K.~Ishimoto, H.~Gad\^{e}lha, and E.~A. Gaffney.
\newblock Filament mechanics in a half-space via regularised {Stokeslet}
  segments.
\newblock {\em Journal of Fluid Mechanics}, 879:808--833, 2019.

\bibitem{arash2020}
Arash Alizad~Banaei, Marco~Edoardo Rosti, and Luca Brandt.
\newblock Numerical study of filament suspensions at finite inertia.
\newblock {\em Journal of Fluid Mechanics}, 882, 2020.

\bibitem{lighthill1975}
M.~J. Lighthill.
\newblock {\em Mathematical biofluiddynamics}.
\newblock Regional conference series in applied mathematics 17. Society for
  Industrial and Applied Mathematics, Philadelphia, 1975.

\end{thebibliography}


\begin{thebibliography}{10}

\bibitem{chorin67}
Alexandre~J. Chorin.
\newblock A numerical method for solving incompressible viscous flow problems.
\newblock {\em J. Comput. Phys.}, 2(1):12--26, 1967.

\bibitem{chorin68}
Alexandre~J. Chorin.
\newblock Numerical solution of the {N}avier--{S}tokes equations.
\newblock {\em Math. Comput.}, 22(104):745--762, 1968.

\bibitem{brown01}
David~L. Brown, Ricardo Cortez, and Michael~L. Minion.
\newblock Accurate projection methods for the incompressible {Navier--Stokes}
  equations.
\newblock {\em J. Comput. Phys.}, 168(2):464--499, 2001.

\bibitem{bell89}
John~B Bell, Phillip Colella, and Harland~M Glaz.
\newblock A second-order projection method for the incompressible
  {Navier--Stokes} equations.
\newblock {\em Journal of Computational Physics}, 85(2):257--283, December
  1989.

\bibitem{almgren96}
Ann~S. Almgren, John~B. Bell, and William~G. Szymczak.
\newblock A numerical method for the incompressible {Navier--Stokes} equations
  based on an approximate projection.
\newblock {\em SIAM J. Sci. Comput.}, 17(2):358--369, 1996.

\bibitem{almgren98}
Ann~S. Almgren, John~B. Bell, Phillip Colella, Louis~H. Howell, and Michael~L.
  Welcome.
\newblock A conservative adaptive projection method for the variable density
  incompressible {Navier--Stokes} equations.
\newblock {\em J. Comput. Phys.}, 142(1):1--46, 1998.

\bibitem{colella90}
Phillip Colella.
\newblock Multidimensional upwind methods for hyperbolic conservation laws.
\newblock {\em J. Comput. Phys.}, 87(1):171--200, 1990.

\bibitem{sussman99}
Mark Sussman, Ann~S. Almgren, John~B. Bell, Phillip Colella, Louis~H. Howell,
  and Michael~L. Welcome.
\newblock An adaptive level set approach for incompressible two-phase flows.
\newblock {\em J. Comput. Phys.}, 148(1):81--124, 1999.

\bibitem{yu03}
Jiun-Der Yu, Shinri Sakai, and James~A. Sethian.
\newblock A coupled level set projection method applied to ink jet simulation.
\newblock {\em {Interfaces and Free Boundaries}}, 5(4):459--482, 2003.

\bibitem{yu07}
Jiun-Der Yu, Shinri Sakai, and James~A. Sethian.
\newblock Two-phase viscoelastic jetting.
\newblock {\em J. Comput. Phys.}, 220(2):568--585, 2007.

\bibitem{rycroft20}
Chris~H Rycroft, Chen-Hung Wu, Yue Yu, and Ken Kamrin.
\newblock Reference map technique for incompressible fluid–structure
  interaction.
\newblock {\em Journal of Fluid Mechanics}, 898, 2020.

\bibitem{kamrin12}
Ken Kamrin, Chris~H. Rycroft, and Jean-Christophe Nave.
\newblock Reference map technique for finite-strain elasticity and fluid--solid
  interaction.
\newblock {\em J. Mech. Phys. Solids}, 60(11):1952--1969, 2012.

\bibitem{valkov15}
Boris Valkov, Chris~H. Rycroft, and Ken Kamrin.
\newblock Eulerian method for multiphase interactions of soft solid bodies in
  fluids.
\newblock {\em J. Appl. Mech.}, 82(4):041011, 04 2015.

\bibitem{sussman94}
Mark Sussman, Peter Smereka, and Stanley Osher.
\newblock A level set approach for computing solutions to incompressible
  two-phase flow.
\newblock {\em J. Comput. Phys.}, 114(1):146--159, 1994.

\bibitem{crank47}
J.~Crank and P.~Nicolson.
\newblock A practical method for numerical evaluation of solutions of partial
  differential equations of the heat-conduction type.
\newblock {\em Mathematical Proceedings of the Cambridge Philosophical
  Society}, 43(1):50--67, 1947.

\bibitem{almgren00}
Ann~S. Almgren, John~B. Bell, and William~Y. Crutchfield.
\newblock Approximate {Projection} {Methods}: {Part} {I}. {Inviscid}
  {Analysis}.
\newblock {\em SIAM J. Sci. Comput.}, 22(4):1139--1159, January 2000.

\bibitem{flory53}
Paul~J. Flory.
\newblock {\em Principles of Polymer Chemistry}.
\newblock Cornell University Press, Ithaca, N. Y., 1953.

\bibitem{richardson11}
Lewis~Fry Richardson.
\newblock {IX}. {T}he approximate arithmetical solution by finite differences
  of physical problems involving differential equations, with an application to
  the stresses in a masonry dam.
\newblock {\em Philos. Trans. Royal Soc. A}, 210(459--470):307--357, 1911.

\bibitem{hairer93}
E.~Hairer, S.~P. N{\o}rsett, and G.~Wanner.
\newblock {\em Solving Ordinary Differential Equations {I}: {N}onstiff
  Problems}.
\newblock Springer, Berlin, 1993.

\bibitem{heath02}
Michael~T. Heath.
\newblock {\em Scientific Computing: An Introductory Survery}.
\newblock McGraw-Hill, 2002.

\bibitem{albensoeder05}
S.~Albensoeder and H.C. Kuhlmann.
\newblock Accurate three-dimensional lid-driven cavity flow.
\newblock {\em Journal of Computational Physics}, 206(2):536--558, July 2005.

\end{thebibliography}
\end{multicols}
\end{document}


\maketitle

\section{Details of numerical algorithms on Cartesian grid}
The numerical algorithms that we use for modeling fluid--structure interactions (FSI) are built upon a modern implementation for simulating incompressible fluid mechanics
that is based upon Chorin's projection method~\cite{chorin67, chorin68}.
Since its introduction in 1968, a variety of extensions have been explored in the literature to improve accuracy and stability~\cite{brown01}.
We make use of a mature implementation that incorporates many such advancements,
developed by Almgren, Bell, and coworkers~\cite{bell89, almgren96, almgren98, colella90, sussman99}.
We refer the reader to the papers by Yu \textit{et al.}~\cite{yu03, yu07} that use this implementation to simulate an inkjet printer nozzle;
these papers provide a comprehensive description of the numerical approaches.

In our algorithms, we keep the fluid simulation component the same as in this existing body of work.
We then build the reference map technique (RMT) on top of the same framework for handling solid objects, using similar principles in the numerical discretization.
The numerical methods and approach are similar to the two-dimensional implementation of the RMT that we previously developed~\cite{rycroft20}.
We use a second-order accurate discretization in space and first-order accurate explicit scheme in time for both the fluid and the solid phases.
Due to error contributions from the interfacial coupling procedure, the overall FSI method is approximately first-order accurate in both space and time.
Convergence rates are reported for various test cases in sections below.

As described in the main text, the treatment of the reference map field and extrapolation is improved and simplified,
and removes the need for specialized level-set methods and reinitialization techniques that have been required in previous implementations~\cite{kamrin12,valkov15,rycroft20}.
A major benefit of the new implementation is that it is more amenable for parallelization in the distributed memory paradigm, a necessity in 3D simulations.

\subsection{Overview}
\label{sub:overall}

\subsubsection*{Governing equations}
We first provide a broad sketch of the overall numerical algorithm and introduce the staggered arrangement of variables on a Cartesian grid.
We discretize the following governing equations for the coupled fluid--structure system,
\begin{align}
    \rho \left (\frac{\p \vu}{\p t}
    + ( \vu \cdot \nabla ) \vu \right) &= - \nabla p +\nabla \cdot \vtau + \vb, \label{eq:full_mmt}\\
    \frac{\p  \vxi}{\p t} + (\vu \cdot \nabla)  \vxi &= \vec{0}, \label{eq:xi_adv}\\
    \nabla \cdot \vu &= 0,
    \label{eq:incomp}
\end{align}
where $\rho$ is the mass density, $\vu$ is the global velocity,
$\vxi$ is the reference map variable in the solid bodies,
$p$ is the global hydrostatic pressure,
$\vec{b}$ is a body force density if applicable,
and $ \vtau$ is the deviatoric stress tensor.
The deviatoric stress tensor $ \vtau$ is the fluid stress tensor $\vtau_f$ in the fully fluid phase, and it is the solid stress tensor $\vtau_s$ in the fully solid phase.
Near the fluid--solid interface, $\vtau$ is a mixed quantity using a smooth transition function over a blur zone of characteristic width $2\epsilon$,
\begin{equation}
  \vtau = H_\epsilon(\phi)  \vtau_f   + (1-H_\epsilon(\phi)) (\vtau_s +  \vtau_a)
  \label{eq:mixed_stresses}
\end{equation}
where $\vtau_a$ denotes the artificial viscous stress tensor added to the solid.
$H_\epsilon(\phi)$ is the smoothed Heaviside function, which is defined in the main text and has been used in other works~\cite{sussman94,yu03,rycroft20}.
To simplify notation, we refer to $\phi_0$, a signed-distance defined in the reference configuration of the solid object, as $\phi$ hereinafter.
The artificial viscous stress is defined as
\begin{equation}
\vtau_a =  \mu_a( 1 + \gamma_t \epsilon H'_\epsilon(\phi)) (\nabla \vu + \nabla \vu^\trans)
\end{equation}
where $\mu_a$ is the solid artificial viscosity and $\gamma_t$ is a dimensionless multiplier to amplify the artificial viscosity in the blur zone.
The contribution of $\nabla \vu^\trans$ term to the divergence of viscous stress is negligible due to the incompressibility constraint $\nabla \cdot \vu = 0$.
Therefore, in the actual computation, we do not compute $\nabla \vu^\trans$ even though it is present in the fluid stress and the artificial stress tensors.

\subsubsection*{Variable arrangement}

\begin{figure}
              \centering
               \includegraphics[width=0.75\textwidth]{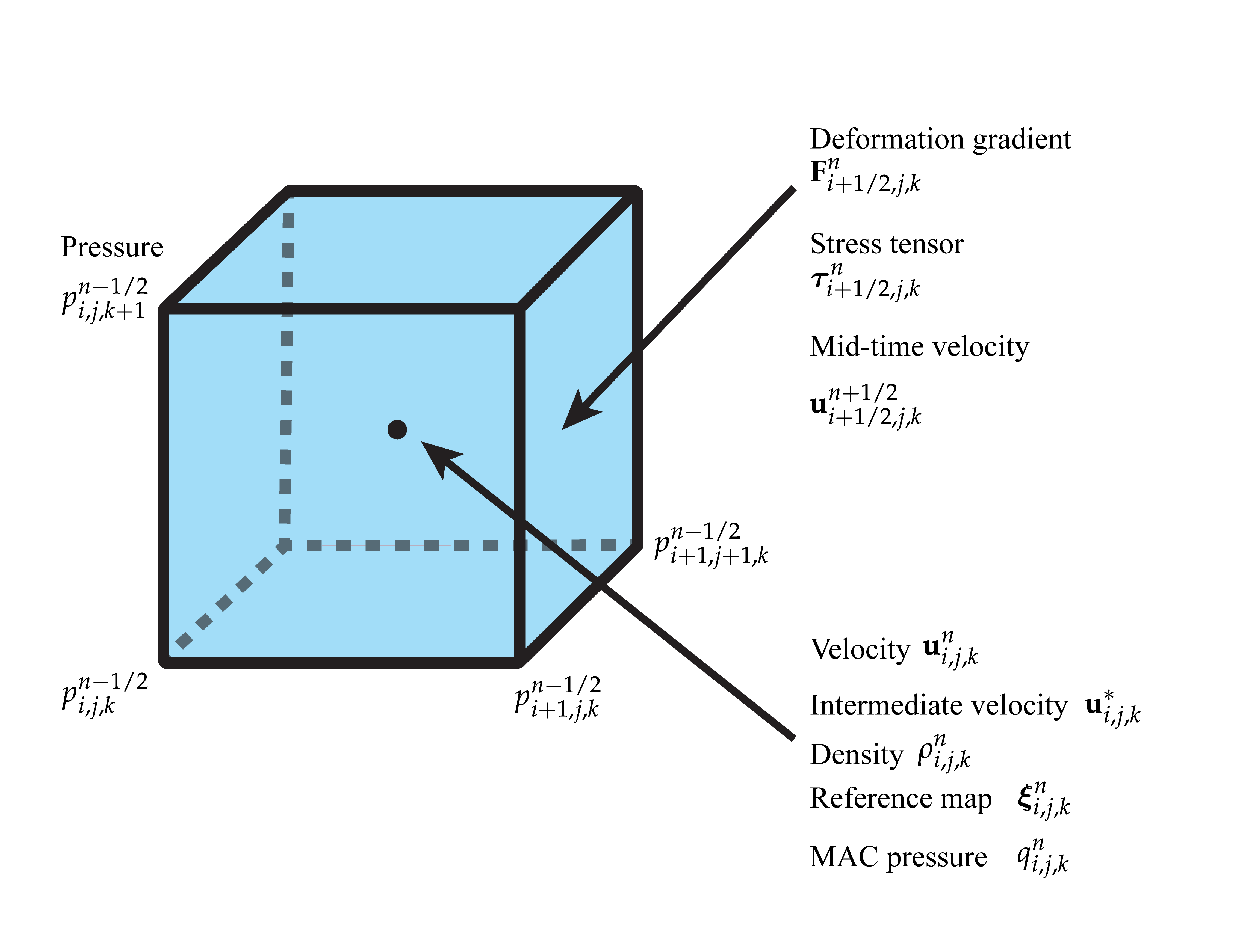}
               \caption{
              Arrangement of variables in the three-dimensional reference map simulation. Depending on the finite difference scheme we use, variable can reside at cell centers, faces, or nodes.
              \label{fig:var_arr}}
            \end{figure}

The placement of variables on the computational grid is illustrated in Fig.~\ref{fig:var_arr}.
A domain $[a_x, b_x] \times [a_y, b_y] \times [a_z, b_z]$ is divided in $M, N, O$ number of grid cells,
yielding grid spacings $\Delta x, \Delta y, \Delta z$, respectively.
To integrate in time, we take steps of size $\Delta t$.
We denote the $n^\text{th}$ timestep $t_n$, and the total number of timesteps $U$.
Thus $t_n = n \Delta t$, $n=1,2,\ldots,U$.
The discretized velocity solution is denoted $\vu_{i,j,k}^n$, where the subscripts $i, j, k$ indicate cell-centered position on the grid,
and the superscript $n$ indicates the number of timesteps.
Other quantities on the grid are indexed in a similar fashion unless specified otherwise.
When applicable, $1/2$ in the indices indicate either cell faces between to grid cells or the midpoint between two timesteps.
When spatial indices are omitted, we refer to the solutions in the entire computation domain.
Primary variables are mass density $\rho_{i,j,k}^n$, global velocity $\vu_{i,j,k}^n$, and reference map variables in solids and blur zones $\vxi_{i,j,k}^n$, which are all placed at the cell centers.
In addition, there is also a global pressure variable \smash{$p_{i,j,k}^{n-1/2}$} placed at the nodes. Note that subscripts $i,j,k$ denote nodal quantities for the pressure field only.

\subsubsection*{Spatial and temporal discretizations}

We discretize Eqs.~\eqref{eq:full_mmt} \& \eqref{eq:xi_adv} by
\begin{align}
\frac{\vu^{n+1} - \vu^n}{\Delta t} + \left[ (\vu \cdot \nabla ) \vu \right]^{n+1/2} &=
\frac{1}{\rho(\phi^{n+1/2})}  \bigg[
- \nabla p^{n+1/2}
+ \nabla \cdot \left ((1- H_\epsilon (\phi^{n+1/2}) )(\vtau_s^{n+1/2} + \vtau_a^{\tilde{n}}) + H_\epsilon (\phi^{n+1/2})\vtau_f^n \right)
+ \vb^{n+1/2} \bigg], \label{eq:dis_full_mmt} \\
\frac{\vxi^{n+1} - \vxi^n} {\Delta t} + \left[ (\vu \cdot \nabla ) \vxi \right]^{n+1/2} &= \vec{0}. \label{eq:dis_xi_adv}
\end{align}
Note that $\phi^{n+1/2}$ refers to $\phi_0(\vxi^{n+1/2})$, not the reinitialized level-set function.
A number of terms on the right hand side (RHS) of the above equations are calculated at the half-timestep $n+\tfrac{1}{2}$ for improved accuracy.
If applicable, a body force density can be prescribed at cell centers at mid timestep, $\mathbf b_{i,j,k}^{n+1/2}$.
Since this is an explicit scheme it is necessary to compute $\vtau_f^n$ and $\vtau_a^{\tilde{n}}$ using information at timestep $n$,
which results in $\vtau_a^{\tilde{n}}$ using quantities at two different times, i.e.,
\begin{equation}
  \vtau_a^{\tilde{n}} = \mu_a( 1 + \gamma_t \epsilon H'_\epsilon(\phi^{n+1/2})) (\nabla \vu^n + (\nabla \vu^n)^\trans).
\end{equation}
Better accuracy could be achieved by using a Crank--Nicolson-type update formula~\cite{crank47}, but since this would depend on $\vu^{n+1}$ it would result in an implicit numerical scheme, which is outside the scope of this work.

To handle the advective terms $ \left[ (\vu \cdot \nabla ) \vu \right]^{n+1/2}$ and $\left[ (\vu \cdot \nabla ) \vxi \right]^{n+1/2}$, we require at cell faces
\begin{enumerate}
  \item intermediate velocities $\vu^{n+1/2}_{i\pm1/2, j,k}, \vu^{n+1/2}_{i,j\pm1/2,k}, \vu^{n+1/2}_{i,j,k\pm1/2}$,
  \item intermediate reference maps $\vxi^{n+1/2}_{i\pm1/2,j,k}, \vxi^{n+1/2}_{i,j\pm1/2,k}, \vxi^{n+1/2}_{i,j,k\pm1/2}$.
\end{enumerate}
We use a Godunov-type upwinding scheme to compute these quantities, which will be discussed in subsection \ref{sub:adv}.
\subsubsection*{MAC projection}
Using the intermediate face velocities, we can compute the discrete divergence in the cell centers using centered finite differences.
However, due to discretization error, this discrete divergence field may not evaluate to precisely zero as expected by Eq.~\eqref{eq:incomp}.
We employ an intermediate marker-and-cell (MAC) projection to correct the face velocities to satisfy the discrete divergence-free property.
This has been shown to improve accuracy and volume conservation in previous work~\cite{rycroft20}.
To do this we first solve the equation
\begin{align}
    \nabla \cdot \left ( \frac{1}{\rho} \nabla q^{n+1/2} \right) = \frac{ \nabla \cdot \vu^{n+1/2}}{\Delta t/2}
    \label{eq:mac_proj}
\end{align}
for an intermediate solution $q^{n+1/2}$ at the cell centers.
The RHS of the equation above is evaluated using the intermediate face velocities.
We subtract $\frac{\Delta t}{2 \rho} \nabla q$ from the intermediate velocities, thus ensuring the discrete divergences are zero to machine precision.
In Eq.~\eqref{eq:mac_proj} we have dropped the subscripts that indicate cell faces, and we will continue to do so for brevity.

\subsubsection*{Approximate pressure-Poisson projection}
Using update equations \eqref{eq:dis_full_mmt}  \& \eqref{eq:dis_xi_adv}, an intermediate velocity update $\vu^{*}$ at $t=t_n$ can be computed.
Once this is computed, we apply the projection step making use of the approximate finite-element projection introduced
by Almgren \textit{et al.}~\cite{almgren96}. This involves solving the Poisson equation
\begin{equation}
  \nabla \cdot \left ( \frac{1}{\rho} \nabla \psi \right) = \frac{\nabla \cdot \left( \vu^{*} - \vu^n \right) }{\Delta t}
\end{equation}
for the pressure correction $\psi$, after which the pressure is updated using $p^{n+1/2} = p^{n-1/2} +  \psi$.
The finite-element projection is implemented using trilinear basis functions centered on each cell corner.
At a particular pressure point $p_{i,j,k}^n$ located at a cell corner, the corresponding basis function
is non-zero over the $2\times 2\times 2$ block of adjacent grid cells.

\subsubsection*{Stress calculations}
The fluid deviatoric stress $\vtau_f^n$ at time $t_n$ is computed at the cell faces by using centered finite differences on the cell centered velocities $\vu^n$.
To compute solid deviatoric stress $\vtau_s^n$, deformation gradient $\vF^n$ is first computed at the cell faces by using centered finite differences on the reference maps, $\vF = (\nabla_x \vxi)^{-1}$.
Then $\vtau_s^n$ is computed at the cell faces directly using the incompressible neo-Hookean constitutive relation
\begin{equation}
  \vtau_s^n = G\left ( \mathbf B^n - \tfrac13\Tr(\mathbf B^n)\right)
\end{equation}
where $\mathbf B^n = \vF^n (\vF^n)^\trans$.
If blur zones of $S$ solid objects overlap, $S>1$,
computations of solid stress and artificial viscous stress remain the same for each object.
To avoid notation confusion, we drop the spatial indices, and use subscript $m$ to represent different solid objects.
But it is assumed that the discussion below only applies at cell faces where it is needed.
In each solid object, we separately compute
\begin{align}
&\vtau_{s, m}^n = G_m\left ( \mathbf B_m^n - \tfrac13\Tr(\mathbf B_m^n)\right),\\
&\vtau_{a, m}^{\tilde{n}} = \mu_a( 1 + \gamma_t \epsilon H'_\epsilon(\phi_m^{n+1/2})) (\nabla \vu^n + (\nabla \vu^n)^\trans),
\end{align}
where $\phi_m^{n+1/2}$ is the specific level-set function defining the fluid--solid interface of solid $m$, evaluated on the reference map variables of this solid object at time $n+1/2$.
In the overlapping region, the overall stress tensors $\vtau_s$ and $\vtau_a$ are combined using the respective stress tensors from each of the overlapping solid objects, 
\begin{equation}
\vtau_s^n =
\begin{cases}
  \sum_{m=1}^S (1-H_\epsilon(\phi_m^n)) \vtau_s^n & \qquad \text{if $\sum_{m=1}^S (1-H_\epsilon(\phi_m^n)) \le 1$,} \\ \\
\frac{\sum_{m=1}^S (1-H_\epsilon(\phi_m^n)) \vtau_s^n } { \sum_{m=1}^S (1-H_\epsilon(\phi_m^n)) } & \qquad \text{otherwise}.
\end{cases}
\end{equation}
The artificial viscous stress tensor  $\vtau_a^n$ is computed similarly.
To blend multiple solid blur zones with the fluid, we modify the volume fraction used in Eq.~\eqref{eq:mixed_stresses} to
\begin{equation}
H_\epsilon(\phi^n) = \max \left[ 1 - \sum_{s=1}^S (1-H_\epsilon(\phi_s^n)), 0 \right].
\label{eq:mod_heaviside}
\end{equation}

\subsection{Evaluating the advective term}
\label{sub:adv}
\subsubsection*{Advective derivatives}
To evaluate advective derivatives on the left hand side of Eqs.~\eqref{eq:dis_full_mmt},~\eqref{eq:dis_xi_adv}, we use the formulae
\begin{align}
\label{eq:adv_terms}
 \left[ (\vu \cdot \nabla ) \vu \right]^{n+1/2} &= \frac{u^{n+1/2}_{i+1/2} + u^{n+1/2}_{i-1/2}}{2} \frac{\vu^{n+1/2}_{i+1/2} - \vu^{n+1/2}_{i-1/2}}{\Delta x} \nonumber \\
 &\phantom{=}+ \frac{v^{n+1/2}_{j+1/2} + v^{n+1/2}_{j-1/2}}{2} \frac{\vu^{n+1/2}_{j+1/2} - \vu^{n+1/2}_{j-1/2}}{\Delta y} \nonumber \\
 &\phantom{=}+  \frac{w^{n+1/2}_{k+1/2} + w^{n+1/2}_{k-1/2}}{2} \frac{\vu^{n+1/2}_{k+1/2} - \vu^{n+1/2}_{k-1/2}}{\Delta z}, \\
  \left[ (\vu \cdot \nabla ) \vxi \right]^{n+1/2}  &= \frac{u^{n+1/2}_{i+1/2} + u^{n+1/2}_{i-1/2}}{2} \frac{\vxi^{n+1/2}_{i+1/2} - \vxi^{n+1/2}_{i-1/2}}{\Delta x} \nonumber \\
  &\phantom{=}+ \frac{v^{n+1/2}_{j+1/2} + v^{n+1/2}_{j-1/2}}{2} \frac{\vxi^{n+1/2}_{j+1/2} - \vxi^{n+1/2}_{j-1/2}}{\Delta y} \nonumber \\
  &\phantom{=}+  \frac{w^{n+1/2}_{k+1/2} + w^{n+1/2}_{k-1/2}}{2} \frac{\vxi^{n+1/2}_{k+1/2} - \vxi^{n+1/2}_{k-1/2}}{\Delta z},
\end{align}
where $\vu=(u,v,w)$. For brevity, only the indices that differ from $i,j,k$ are shown in the subscripts. To obtain $\vu^{n+1/2}$ and $\vxi^{n+1/2}$ at a cell face, we use a Taylor expansion using values straddling the cell face, keeping terms up to first order in time and space. Without loss of generality, we consider a face in $x$-direction, indexed by ${i+1/2, j, k}$. Two Taylor expansions are constructed, centered at values at cell $i,j,k$ and cell ${i+1,j,k}$.
Expanding from the left side of the face, we have
\begin{align}
\vu^{n+1/2,\text{L}}_{i+1/2,j,k} = \vu^{n}_{i,j,k} + \frac{\Delta x}{2} \frac{\p \vu}{\p x} \bigg |^n_{i,j,k} + \frac{\Delta t}{2} \frac{\p \vu}{\p t} \bigg |^n_{i,j,k}.
\label{eq:taylor_left}
\end{align}
From the momentum balance equation (Eq.~\eqref{eq:full_mmt}) we obtain an expression for $\p \vu/\p t$ as
\begin{align}
\frac{\p \vu}{\p t}\bigg |^n_{i,j,k}  = - \left ( u \frac{\p \vu^n}{\p x}
+ \overline{\left(v\frac{\p \vu^n}{\p y}\right)}
+ \overline{\left(w\frac{\p \vu^n}{\p z}\right)} \right )_{i,j,k}
+ \frac{1}{\rho(\phi^n)_{i,j,k}} \left (- \nabla p^{n-1/2} + \nabla \cdot \vtau^n(\vu^n, \vxi^n, \phi^n) + \vb^n \right)_{i,j,k}.
\label {eq:time_partial}
\end{align}
The terms on the RHS of Eq.~\eqref{eq:time_partial} are evaluated at step $n$, except for
the pressure, which is evaluated at step $n-1/2$.
Since we are expanding in the $x$-direction, the barred terms in the bracket
are the tangential derivatives (also called transverse derivatives) and they
are treated differently; details will be discussed later in this subsection.
Substituting Eq.~\eqref{eq:time_partial} into Eq.~\eqref{eq:taylor_left}, we have
\begin{align}
  \vu^{n+1/2,\text{L}}_{i+1/2,j,k} &= \vu^{n}_{i,j,k} + \left( \frac{\Delta x}{2} - \frac{\Delta t}{2} u^n_{i,j,k} \right) \frac{\p \vu}{\p x} \bigg |^n_{i,j,k}
                                   - \frac{\Delta t}{2} \left (\overline{\left(v\frac{\p \vu^n}{\p y}\right)} + \overline{\left(w\frac{\p \vu^n}{\p z}\right)} \right )_{i,j,k} \nonumber \\
                                   &\phantom{=} +  \frac{\Delta t }{2 \rho(\phi^n)_{i,j,k}} \left (- \nabla p^{n-1/2} + \nabla \cdot \vtau^n(\vu^n, \vxi^n, \phi^n) + \vb^n \right)_{i,j,k}.
\label{eq:left_final}
\end{align}
Similarly, Taylor expanding from the right side of the cell face, we have
\begin{align}
  \vu^{n+1/2,\text{R}}_{i+1/2,j,k} &= \vu^{n}_{i+1,j,k} - \left( \frac{\Delta x}{2} + \frac{\Delta t}{2} u^n_{i+1,j,k} \right) \frac{\p \vu}{\p x} \bigg |^n_{i+1,j,k} 
                                   - \frac{\Delta t}{2}\left (\overline{\left(v\frac{\p \vu^n}{\p y}\right)} + \overline{\left(w\frac{\p \vu^n}{\p z}\right)} \right )_{i+1,j,k} \nonumber \\
  & \phantom{=}+ \frac{\Delta t}{2\rho(\phi^n)_{i+1,j,k}} \left (- \nabla p^{n-1/2} + \nabla \cdot \vtau^n(\vu^n, \vxi^n, \phi^n) + \vb^n \right)_{i+1,j,k}.
\label{eq:right_final}
\end{align}
The Taylor expansion procedure to create values of $\vxi^{n+1/2}_{i+1/2, j, k}$ makes use of Eq.~\eqref{eq:xi_adv}, which has a much simpler form. Expanding from the left and right sides respectively gives
\begin{align}
\vxi^{n+1/2,\text{L}}_{i+1/2,j,k} =& \vxi^{n}_{i,j,k} + \left( \frac{\Delta x}{2} - \frac{\Delta t}{2} u^n_{i,j,k} \right) \frac{\p \vxi}{\p x} \bigg |^n_{i,j,k}
- \frac{\Delta t}{2} \left (\overline{\left(v\frac{\p \vxi^n}{\p y}\right)} + \overline{\left(w\frac{\p \vxi^n}{\p z}\right)} \right )_{i,j,k},  \\
\vxi^{n+1/2,\text{R}}_{i+1/2,j,k} =& \vxi^{n}_{i+1,j,k} - \left( \frac{\Delta x}{2} + \frac{\Delta t}{2} u^n_{i+1,j,k} \right) \frac{\p \vxi}{\p x} \bigg |^n_{i+1,j,k}
- \frac{\Delta t}{2} \left (\overline{\left(v\frac{\p \vxi^n}{\p y}\right)} + \overline{\left(w\frac{\p \vxi^n}{\p z}\right)} \right )_{i+1,j,k}.
\label{eq:xi_taylor_final}
\end{align}

\subsubsection*{Godunov-type upwinding scheme}
\label{sub:godunov}
Now there are two choices for velocity and two for reference map on the cell face indexed by ${i+1/2,j,k}$, we perform Godunov upwinding to select one.
We define the normal velocity as $u_{\nor}^{n+1/2, \text{L}}$ and $u_{\nor}^{n+1/2, \text{R}}$, expanded from the left and the right cells, respectively.
We use $q^{n+1/2, \text{L}}$ and $q^{n+1/2, \text{R}}$ to represent other components of the velocity or the reference map, expanded from the left and the right, respectively.
To choose the normal velocity, we follow
\begin{align}
u_{\nor}^{n+1/2} =
\begin{cases}
u_{\nor}^{n+1/2, \text{L}} & \text{if~} u_{\nor}^{n+1/2, \text{L}}  > 0 \text{~and~} u_{\nor}^{n+1/2, \text{L}}  + u_{\nor}^{n+1/2, \text{R}}  > 0, \\
u_{\nor}^{n+1/2, \text{R}} & \text{if~} u_{\nor}^{n+1/2, \text{R}} < 0 \text{~and~} u_{\nor}^{n+1/2, \text{L}}  + u_{\nor}^{n+1/2, \text{R}}  < 0, \\
0  & \text{otherwise.}
\end{cases}
\end{align}
After this, we select a value for the other quantities at the cell face based on $u_{\nor}^{n+1/2}$ according to
\begin{align}
q^{n+1/2} =
\begin{cases}
q^{n+1/2, \text{L}} & \text{if~} u_{\nor}^{n+1/2}  > 0, \\
\frac{q^{n+1/2, \text{L}}  + q^{n+1/2, \text{R}} }{2}  & \text{if~} u_{\nor}^{n+1/2} = 0, \\
q^{n+1/2, \text{R}} & \text{if~} u_{\nor}^{n+1/2} < 0.
\end{cases}
\end{align}
Here we have dropped all spatial index subscripts since all terms are evaluated at $i+1/2, j, k$.

\subsubsection*{Normal derivatives}
\label{sub:norm}

To compute the normal derivatives (in the example given here, $\p \vu/\p x$ and  $\p \vxi/\p x$), we employ the fourth-order monotonicity-limited scheme of Collela~\cite{colella90}, which is described in detail by Yu \textit{et al.}~\cite{yu03} and Rycroft \textit{et al.}~\cite{rycroft20}.

\subsubsection*{Tangential derivatives}
\label{sub:tang}

To ensure stability, especially in intermediate to high Reynolds number regime, we compute the tangential derivatives (shown as the barred terms in Eqs.~\eqref{eq:left_final}, \eqref{eq:right_final}, \& \eqref{eq:xi_taylor_final}) using an upwinding scheme which is commonly used for solving hyperbolic conservative laws ~\cite{bell89,almgren96,sussman99,yu03}.
Our approach to construct an upwinding scheme in 3D is only one of the many possibilities, and in selecting the following scheme, we prioritize algorithmic simplicity and the ease implementation.

Without loss of generality, we consider the quantities on the transverse faces indexed by ${i,j+1/2,k}$, required in Eq.~\eqref{eq:left_final},~\eqref{eq:right_final},~\eqref{eq:xi_taylor_final} when applied to faces indexed by ${i+1/2,j,k}$.
The procedure to compute tangential derivative terms along the $z$-direction is similar.
We start by performing a Taylor expansion again to construct the velocities and reference maps on the transverse cell faces, but neglect contributions from pressure, stress, body forces, and tangential derivatives to obtain
\begin{align}
\bar \vu^{n+1/2,\text{D}}_{i,j+1/2,k} =& \vu^{n}_{i,j,k} + \left( \frac{\Delta y}{2} - \frac{\Delta t}{2} v^n_{i,j,k} \right) \frac{\p \vu}{\p y} \bigg |^n_{i,j,k}, \\
\bar \vu^{n+1/2,\text{U}}_{i,j+1/2,k} =& \vu^{n}_{i,j+1,k} - \left( \frac{\Delta y}{2} + \frac{\Delta t}{2} v^n_{i,j+1,k} \right) \frac{\p \vu}{\p y} \bigg |^n_{i,j+1,k}, \\
\bar \vxi^{n+1/2,\text{D}}_{i,j+1/2,k} =& \vxi^{n}_{i,j,k} + \left( \frac{\Delta y}{2} - \frac{\Delta t}{2} v^n_{i,j,k} \right) \frac{\p \vxi}{\p y} \bigg |^n_{i,j,k}, \\
\bar \vxi^{n+1/2,\text{U}}_{i,j+1/2,k} =& \vxi^{n}_{i,j+1,k} - \left( \frac{\Delta y}{2} + \frac{\Delta t}{2} v^n_{i,j+1,k} \right) \frac{\p \vxi}{\p y} \bigg |^n_{i,j+1,k},
\end{align}
where the superscripts D and U denote down and up, respectively.
We perform a similar Godunov upwinding procedure to select a normal advective velocity at each face, so that
\begin{align}
\bar v_{\text{adv}}^{n+1/2} =
\begin{cases}
\bar v^{n+1/2, \text{D}} & \text{if~} \bar v^{n+1/2, \text{D}} > 0 \text{~and~}  \bar v^{n+1/2, \text{D}} +  \bar v^{n+1/2, \text{U}} > 0, \\
\bar v^{n+1/2, \text{U}} & \text{if~} \bar v^{n+1/2, \text{U}} < 0 \text{~and~}  \bar v^{n+1/2, \text{D}} +  \bar v^{n+1/2, \text{U}} < 0, \\
0  & \text{otherwise,}
\end{cases}
\end{align}
where we have dropped the subscripts since all terms are evaluated at $i,j+1/2,k$.
Next, we select between Taylor expansions $\bar \vu^{n+1/2,\text{D}}$ and $\bar \vu^{n+1/2,\text{U}}$, as well as between expansions $\bar \vxi^{n+1/2,\text{D}}$ and $\bar \vxi^{n+1/2,\text{U}}$, based on $\bar v_{\text{adv}}^{n+1/2}$.
For two generic vector quantities at the face denoted by $\bar \vq^{n+1/2,\text{D}}$ and $\bar \vq^{n+1/2,\text{U}}$, we define
\begin{align}
\bar \vq^{n+1/2} =
\begin{cases}
\bar \vq^{n+1/2, \text{D}} & \text{if~} \bar v_\text{adv}^{n+1/2} > 0, \\
\left ( \bar \vq^{n+1/2, \text{D}} + \bar \vq^{n+1/2, \text{U}} \right) / 2 & \text{if~} \bar v_\text{adv}^{n+1/2} = 0, \\
\bar \vq^{n+1/2, \text{U}} & \text{if~} \bar v_\text{adv}^{n+1/2} < 0.
\end{cases}
\end{align}
Finally, we compute the tangential derivative terms,
\begin{align}
\overline {\left( v\frac{\p \vq}{\p y} \right) }\bigg |^{n+1/2} _{i,j,k}
= \frac{\bar v^{n+1/2}_{i,j-1/2, k,\text{adv}} + \bar v^{n+1/2}_{i,j+1/2,k, \text{adv}}}{2} \frac{\bar \vq^{n+1/2}_{i,j+1/2,k} - \bar \vq^{n+1/2}_{i,j-1/2,k}}{\Delta y}.
\end{align}

\subsection{Contact with a wall}
We use subscript $\alpha$ to represent walls of the rectangular domain, with $\alpha=1,2,3,4,5,6$ referring to walls with normal vectors $\vec{e}_x, -\vec{e}_x, \vec{e}_y, -\vec{e}_y, \vec{e}_z, -\vec{e}_z$, respectively.
The inward unit normal at a wall is denoted by $\nor_\alpha$.
When a part of the solid body is within a threshold distance from a wall with $\nor_\alpha$, we impose a repulsive acceleration, $\mathbf a_{\text{rep},\alpha}$, to the grid cells in the solid body that have breached the threshold, acting at the cell centers.
This acceleration is multiplied by a transition function in a small transition zone of width $\Delta h_\alpha$, where $\Delta h_\alpha$ is the grid spacing in the Cartesian direction aligned with $\nor_\alpha$.

For those solid grid cells that experience wall acceleration,
\begin{equation}
  \mathbf a_{\text{rep}, \alpha} = \left\{
  \begin{array}{ll}
    A_\text{rep}  \nor_\alpha & \qquad \text{if $\phi\le-\Delta h_\alpha$,} \\
    \left ( \frac{1}{2}- \frac{\phi}{2\Delta h_\alpha} \right) A_\text{rep}  \nor_\alpha & \qquad \text{if $|\phi|<\Delta h_\alpha$,} \\
    0 & \qquad \text{if $\phi\ge\Delta h_\alpha$}.
  \end{array}
  \right.
  \label{eq:wall_acc}
\end{equation}
If a solid body is in close approach to more than one wall, the repulsive accelerations are added together.
As the solid approaches the wall and deforms to form a contact area, a small region of cells experiences the repulsive acceleration described by Eq.~\eqref{eq:wall_acc}.

We use the following approach to choose an appropriate value for $A_\text{rep}$.
First, given the velocity of an object at its center of mass, the repulsive acceleration should be sufficient to stop the object from going through any physical boundary.
In a rigid body, this means that the center of mass velocity is zero before or when the edge of the solid reaches the boundary.
Though the solid objects will deform in our simulations, the rigid body case offers good guiding intuition.
Second, we impose the condition that the repulsive acceleration, at a minimum, should be able to support an object resting against a wall.

We first define a critical threshold, $d_w$. When the distance between any cell belonging to the solid object and the wall falls below $d_w$, the wall repulsive acceleration comes into effect.
In practice, we set $d_w = \Delta h_\alpha$ so that the solid objects are sufficiently close to the wall and experience effects due to boundaries,
however, $d_w$ can be increased so that the repulsive acceleration can be applied to more grid cells.

Let us consider an object of volume $V$ approaches the wall with a normal velocity component $U$, which has a momentum along the normal direction, $\rho_s V U$.
Suppose the volume of the object that actually experiences repulsive acceleration is $V_\text{rep}$,
the effective acceleration at the center of mass is $\frac{V_\text{rep} A_\text{rep} }{ V}$.
Following the reasoning of the first case above,
\begin{equation}
\begin{aligned}
&U  - \frac{V_\text{rep} A_\text{rep} }{ V} t_s = 0, \\
&U t_s - \frac12 \frac{V_\text{rep} A_\text{rep} }{ V} t_s^2 = d_w.
\end{aligned}
\end{equation}
Solving these equations yields $A_\text{rep} = \frac{V U^2}{2d_wV_\text{rep}}$ and $t_s = 2d_w/U$.

We use the smallest length scale of the solid, $L_\text{min}$,  to obtain a conservative estimate of the contact area between the solid and the wall, and use $d_w$ as the thickness of the region of cells that breach this threshold.
Thus a conservative estimate of $V_\text{rep}$ is $V_\text{rep} = d_w L^2_\text{min}$, which yields
\begin{equation}
A_\text{rep} =\frac{V U^2}{2d_w^2 L^2_\text{min}}.
\label{eq:rep1}
\end{equation}
In the second case described above, when the object is at rest at a wall, we need to make sure that the forces experienced by the region of cells that have breached the $d_w$ threshold is large enough to support the entire object, so that $|\rho_s - \rho_f| Vg = \rho_s V_\text{rep} A_\text{rep}$ and hence
\begin{equation}
	A_\text{rep} = \frac{|\rho_s - \rho_f| Vg}{ \rho_s V_\text{rep}}.
	\label{eq:rep2}
\end{equation}
In practice, we also find that it is helpful to set a minimum value of $A_\text{re} = U^2_\text{safe}/d_w$, $U_\text{safe} \approx 4.5$, as an additional safety measure to prevent solid objects from moving through physical boundaries. Given this lower limit and Eqs.~\eqref{eq:rep1},~\eqref{eq:rep2}, we choose the maximum value among the three to be the value of $A_\text{rep}$,
\begin{equation}
 A_\text{rep} = \max \left( U^2_\text{safe}/d_w, \frac{V U^2}{2d_w^2 L^2_\text{min}}, \frac{|\rho_s - \rho_f| Vg}{ \rho_s V_\text{rep}} \right).
\end{equation}
For simulations with multiple objects, $A_\text{rep}$ is computed for each object separately.

\subsection{Weighting scheme for weighted least squares-based extrapolation}
In the absence of the reinitialized level-set function, we use an exponential decay kernel centered at the extrapolated cell to weigh data points in the linear squares regression.
Near the fluid--solid interface we incorporate approximate geometric information via $\phi_0(\vxi)$.

Consider a cell $(p,q,r)$ which provides data $\vxi_d$ at position $\vx_d=(x_p, x_q, x_r)$ in the weighted least squares linear regression to extrapolate
to $\vxi_e$ at cell $(i,j,k)$ with position $\vx_e = (x_i, x_j, x_k)$.
First, we compute the normalized gradient vector $\nor_e$ of $\phi_0(\vxi^n)$ at cell $(i,j,k)$
as well as a physical vector $\vx' = \vx_e - \vx_d$.
Then, depending on which layer extrapolation $\vxi_e$ resides in, the weight of data provided by cell $(p,q,r)$ is defined as
\begin{equation}
\label{eq:weight}
\omega=
\left \{
  \begin{array}{ll}
   \max\left( 0, \frac{\vx' \cdot \nor_e}{|\vx'|} {2 ^{- (r_i + r_j + r_k)}} \right) & \quad \text{if $l_{i, j, k}\le2$, } \\
    {2 ^{- (r_i + r_j + r_k)}} &  \quad \text{if $l_{i, j, k}>2$,}
  \end{array}
  \right.
\end{equation}
where $r_i = |i-p|, r_j=|j-q|, r_k = |k-r|$ and $l_{p,q,r}$ is the index of the extrapolation layer as defined in the main text.
The exponential kernel accounts for locality on the Cartesian grid,
while $\nor_e$ approximates a surface normal and accounts for geometric information near the interface.
However, we only make use of $\nor_e$ in the extrapolating to the first two layers,
since $\phi (\vxi^n)$ approximates a signed-distance function more poorly farther away from the interface in the presence of deformation.

\subsection{Configuring a simulation}
\label{sub:setup}
Before commencing a simulation, we need expressions of level-set functions that represent the fluid--solid interface.
Each solid body is associated with such as level-set function.
In addition, we need to configure the simulation domain, and prescribe material properties, boundary conditions, and initial conditions.
Besides physics-related parameters, we  also need to specify the type of data to collect and the frequency of outputting data files and checkpoint files.
Main parameters that specify the physics in a simulation are documented in Table~\ref{table:pars}.

\subsubsection*{Initialization}
\begin{enumerate}
    \item Set initial conditions for velocity $\vu_{i,j,k}^0$, and for reference map $\vxi_{i,j,k}^0$ for all $i, j, k$ such that $\phi_0(\vxi_{i,j,k}^0) \le 0$.
    We refer to these reference map variables as primary reference maps (PRMs).
    Using the undeformed state as the reference state of a solid body, the primary reference map variable is set to be the physical coordinate on the grid, $\vxi_{i,j,k}^0 = \vx_{i,j,k}$, where $\vx=(x_i, y_j, z_k)$.

    \item If a test case has known pressure, also set $p_{i,j,k}^{-1/2}$, otherwise, we may set pressure to 0 everywhere.
    However, since pressure is computed as an auxiliary field rather than from an equation of state, the initial pressure profile is not known in general.
    Therefore, $p_{i,j,k}^{-1/2}=0$ is generally not compatible with the initial velocity.
    The incompatibility can degrade the accuracy of the solution.

    \qquad To address this, we perform initial iterations to estimate $p_{i,j,k}^{-1/2}$. We run the timestepping algorithm, described in subsection \ref{sub:timestepping}, with an initial guess of $p_{i,j,k}^{-1/2}=0$.
    After an iteration, we keep the pressure estimate $\tilde p_{i,j,k}^{1/2}$ but discard changes to all other fields.
    In the next iteration, let $p_{i,j,k}^{-1/2} = \tilde p_{i,j,k}^{1/2}$, and repeat until a tolerance on the divergence of $\vu_{i,j,k}^0$ is reached, or until a maximum number of iterations is reached.
    Note that in order to carry out initial iterations, the variant of approximate projection method that computes the pressure increments should be used \cite{almgren00}.

    \item If applicable, initialize $\mathbf b_{i,j,k}^0$.
    \item Extrapolate reference map variables to the blur zones.
    Extrapolated reference maps (ERMs) and the PRMs (together denoted as  $\vxi_{i,j,k}^0$) are both required to begin the simulation.
    \item Evaluate $\phi_0(\vxi^0)$ everywhere reference map variables  $\vxi_{i,j,k}^0$ exist.
    \item Initialize $\rho_{i,j,k}^0$ by blending $\rho_f$ and $\rho_s$ using using $H_\epsilon(\phi_0(  \vxi_{i,j,k}^0 ))$ (Eq.~\eqref{eq:mod_heaviside}).
\end{enumerate}

%
\subsection{Timestepping algorithm}
\label{sub:timestepping}

After initializing the simulation, we perform the following steps for timestep $n=0, 1, 2, \ldots, U$ of size $\Delta t$.
This procedure is also used in the iterations to create an initial pressure profile.
\begin{enumerate}
	\item Compute $\vtau_f^n$, $\vtau_s^n$, and $\vtau_a^n$, and mix them to get $\vtau^n$, as described in subsection \ref{sub:overall}.

    \item Using the Godunov-type upwinding scheme described in subsection \ref{sub:godunov}, we compute $\vu_{i\pm1/2,j,k}^{n+1/2}, \vu_{i,j\pm1/2,k}^{n+1/2}, \vu_{i,j,k\pm1/2}^{n+1/2}$ everywhere, and $\vxi_{i\pm1/2,j,k}^{n+1/2}, \vxi_{i,j\pm1/2,k}^{n+1/2}, \vxi_{i,j,k\pm1/2}^{n+1/2}$ only within solids.

	\item Perform the MAC projection, then update the normal face velocities.
	Any boundary conditions should be enforced before the projection step and after the velocity update.

	\item Compute the advective term $ [(\vu \cdot \nabla) \vu ]_{i,j,k}^{n+1/2}$ and $ [(\vu \cdot \nabla) \vxi ]_{i,j,k}^{n+1/2}$ according to Eqs.~\eqref{eq:adv_terms}.

	\item Compute the predictor values of PRMs, i.e. $\vxi_{i,j,k}^{n+1}$ within solids (Eq.~\eqref{eq:dis_xi_adv}).

	\item Extrapolate PRMs to yield ERMs to cover the blur zones, these are the predictor values of ERMs.

	\item Apply corrector to all available reference maps (PRMs and ERMs), $\vxi_{i,j,k}^{n+1/2} = \frac12 \left(\vxi_{i,j,k}^{n} + \vxi_{i,j,k}^{n+1}\right) $.

	\item Compute solid stress  $\vtau_s^{n+1/2}$ and artificial viscous stress $\vtau_a^{\tilde{n}}$, then mix with $\vtau_f^n$ to create $\vtau^{n+1/2}$, as described in subsection \ref{sub:overall}.

	\item Update density to $\rho_{i,j,k}^{n+1/2}$ using $H_\epsilon(\phi_0(  \vxi_{i,j,k}^{n+1/2} ))$ (Eq.~\eqref{eq:mod_heaviside}).

	\item Compute the intermediate velocity for approximate projection, $\vu_{i,j,k}^{*}$, by
	\begin{equation} \vu_{i,j,k}^{*} = \vu_{i,j,k}^n - \Delta t [(\vu \cdot \nabla) \vu ]_{i,j,k}^{n+1/2} - \frac{1}{\rho_{i,j,k}^{n+1/2}} \left( \nabla \cdot  \vtau^{n+1/2} - \nabla p_{i,j,k}^{n-1/2} + \mathbf b_{i,j,k}^{n+1/2}\right)
	\end{equation}

	\item Perform the approximate projection, then update cell center velocities to $\vu_{i,j,k}^{n+1}$ and nodal pressure to $p_{i,j,k}^{n+1/2}$. Any boundary conditions should be enforced before the projection step and after updating velocity and pressure.

	\item Restore all PRMs and ERMs to their predictor values, $\vxi_{i,j,k}^{n+1}$.

	\item Update the density to $\rho_{i,j,k}^{n+1}$ using $H_\epsilon(\phi_0(  \vxi_{i,j,k}^{n+1/2} ))$ (Eq.~\eqref{eq:mod_heaviside}).
\end{enumerate}

\section{Test cases}
    In this section, we first present simulations of a pre-stretched, immersed sphere to demonstrate that
    the convergence of our method, as well as the scaling behavior of the simulated FSI problem, are as expected.
    We also provide additional details and results of examples mentioned in the main text.

	\subsection{Pre-stretched sphere}  We simulate a viscoelastic neo-Hookean solid sphere, subject to initial strain, relaxing to equilibrium in a cubic periodic domain.
	Parameters reported here are dimensionless.

\begin{figure}
            \centering
                \includegraphics[width=\textwidth]{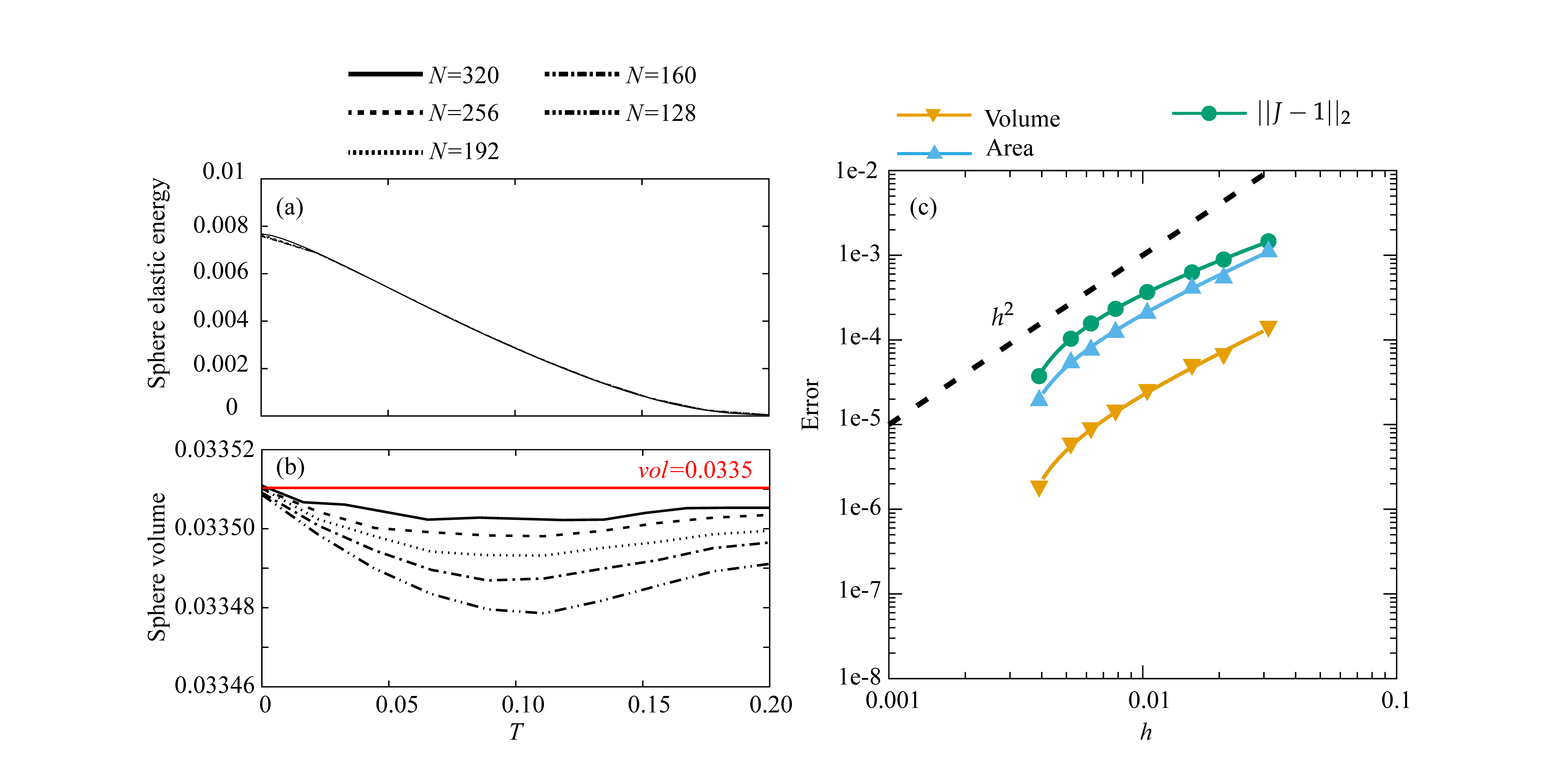}
                \caption{ (a) Elastic energy and volume of a sphere with radius $0.2$ and shear modulus $G=1$ in a periodic unit box.
                The incompressible sphere is stretched in $x$-direction by a factor of $1.44$, and compressed in the other directions accordingly.
                 Fluid viscosity $\mu=0.01$, and solid sphere has the same density as the fluid, $\rho_s=\rho_f=1$.
                 An incompressible neo-Hookean solid is modeled.
                 However,  $J \ne 1$ in our numerical scheme, especially near the fluid--solid interface. Therefore, a correction factor is used in the strain energy density (Eq.~\eqref{eq:elas}).
                 The correction approaches zero as the grid is refined, since $J$ converges to 1.
                (b) Convergence of errors in surface area, volume, elastic energy, and $\|J-1\|_2$, where $J$ is the determinant of the deformation gradient.
                Errors are measured at $T=0.2$.
                The solid lines are fitted curves using Eq.~\eqref{eq:3params}.
                Convergence rates (with Richardson correction factor $\alpha$ in brackets) are $1.34 (0.97), 1.44 (1.0), \text{~and~} 1.04 (1.0)$, for area, volume, and $\|J-1\|_2$, respectively.}
                \label{fig:str_conv}
            \end{figure}

\begin{figure}
            \centering
                \includegraphics[width=\textwidth]{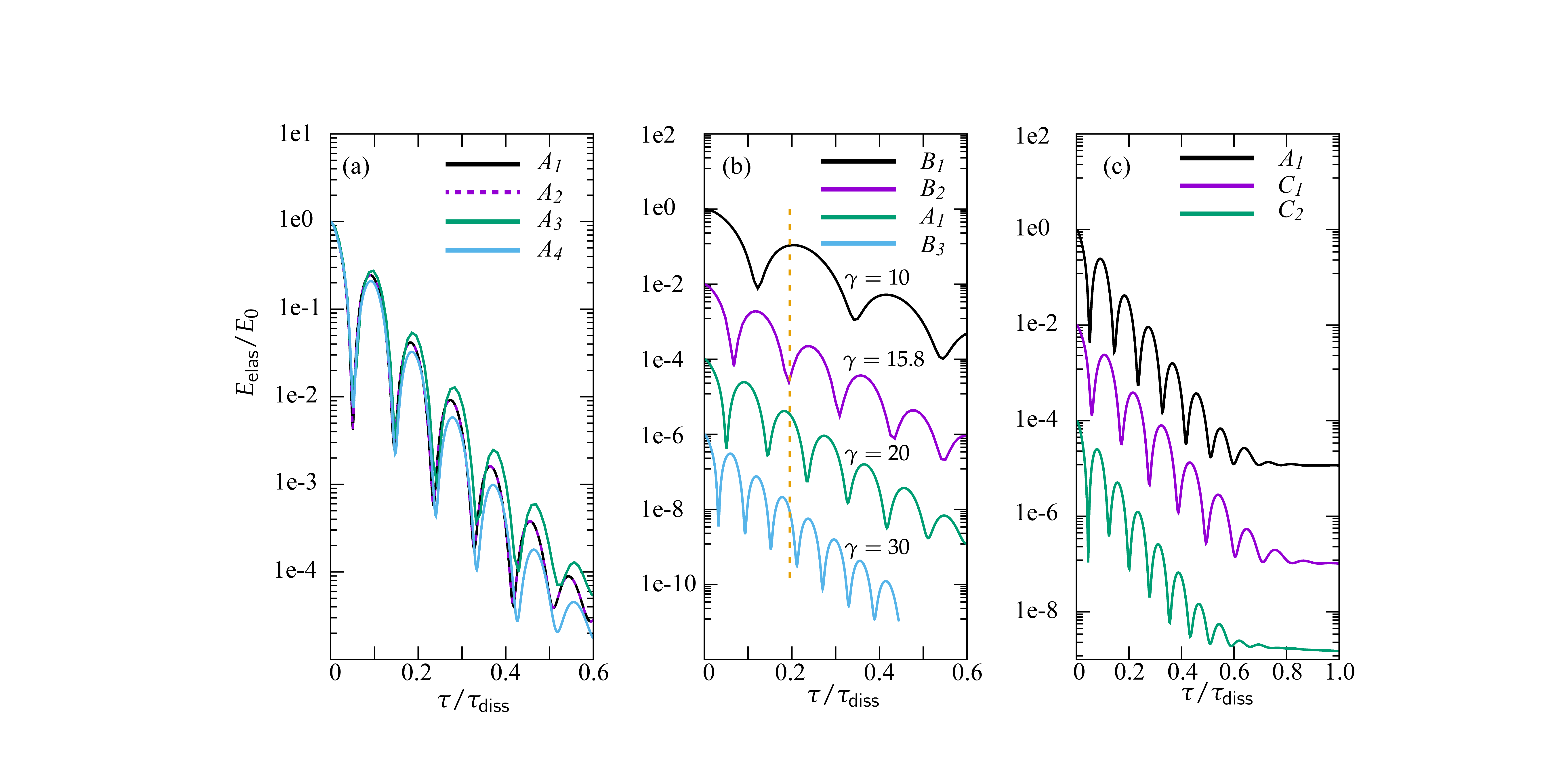}
                \caption{ Elastic energy (rescaled by respective initial value $E_0$) over time (rescaled by respective characteristic time $\tau_\text{diss}$) of pre-stretched spheres.
                 Simulation parameters are documented in Table~\ref{table:str_params}.
                 In (b), normalized elastic energies are further reduced by a factor of $100$, $10^4$, $10^6$ for simulation $B_2$, $A_1$, $B_3$
                 so that the curves are clearly separated.
                 In (c), similar reduction factors of $100$ and $10^4$ are applied to curves from simulation $C_1$ and $C_2$ for clarity.
                }
                \label{fig:str_scaling}
            \end{figure}

	\begin{enumerate}
		\item The sphere has a radius $R=0.2$ and a shear modulus $G=1$, positioned at the center of a periodic box with unit length, fluid viscosity $\mu = 0.01$.
		The sphere has the same density as the fluid, and is pre-strained at $T=0$ in the $x$-direction by a stretch $\lambda=1.21$.
		We use isotropic grid spacing $\Delta x = \Delta y = \Delta z = h$.
		The solution with the smallest $h$ is used as the reference solution in the convergence plot.

		\qquad Elastic energy and volume of the sphere over time are reported in Fig.~\ref{fig:str_conv}(a),
		and the convergence in surface area, volume, and $\|J-1\|_2$ is reported in Fig.~\ref{fig:str_conv}(b), where
		$J$ is the determinant of the deformation gradient.
		The pressure-Poisson projection method we use to enforce incompressibility is an approximate one,
		which means locally, it is possible to have $J \ne 1$.
        To account for the local violation of incompressibility, we compute strain energy density with a logarithmic correction factor \cite{flory53},
		\begin{align}
		\label{eq:elas}
		\Phi_\text{elas} = \frac12 G\big (\Tr(\vC) - 3 - 2\ln(J) \big)
		\end{align}
		where $\vC=\vF^\trans \vF$ is the right Cauchy--Green tensor.
		This strain energy density equation is used throughout our analysis.

		\qquad Since we use a reference solution rather than the exact solution,
		only part of the error can be captured by a Richardson error model \cite{richardson11, hairer93, heath02, rycroft20}.
		For other sources of error, e.g.~reference map extrapolations, Godunov-type upwinding procedures, Richardson model is not a good fit.
		Thus, we adopt the 3-parameter error model proposed by Rycroft \textit{et al.}~\cite{rycroft20}
		\begin{align}
		\label{eq:3params}
		E(h) = B(h^s - \alpha h_\text{ref}^s) + \mathcal{O}(h^{s+1})
		\end{align}
		where $\alpha$ is the Richardson correction factor which indicates the proportion in the error that can be captured by a Richardson error model, and $s$ is the overall convergence rate.
		We report the fitted convergence parameters in the caption of Fig.~\ref{fig:str_conv}.

		\item We demonstrate scaling of elastic energy of the immersed sphere in various combinations of $G$, $\mu$, $\rho_s/\rho_f$, and $R$ (Fig.~\ref{fig:str_scaling}). First, we find the pertinent parameter in our test case by dimensional analysis.
		Consider a standalone, ideal, incompressible neo-Hookean elastic body with density $\rho_s=\rho$, shear modulus $G$, and a characteristic length scale $R$.
		In the absence of viscous damping, uniaxial tension or compression induces an oscillatory response due to elastic restoring forces in the solid body.
		The oscillation time scale is $\tau_\text{osc} = R\sqrt{\rho/G}$.
		In addition to being the characteristic oscillation period, this time scale can also be interpreted as the time it takes for elastic waves to traverse the elastic body.
		Now, suppose the elastic body is immersed in a viscous fluid, with viscosity $\mu$ and density $\rho_f = \rho$.
		In our method, the solid material is made viscoelastic by using a  constant artificial viscosity $\mu_a$.
		We impose $\mu_a = \mu$ in this case.
		Out of the parameters relevant to the fluid $\rho, \mu, R$, we can build another time scale, $\tau_\text{diss} = \rho R^2 / \mu$.
		We can interpret $\tau_\text{diss}$ as the time scale of the fluid dissipating the kinetic energy generated by the oscillating sphere as it returns to equilibrium.

		\qquad The ratio between these two time scales is
		$$\gamma = \tau_\text{diss} / \tau_\text{osc} = {R \sqrt{\rho G}}/{\mu}.$$
		In the simple system we have considered here, having assumed solid and fluid have identical densities $\rho$, and viscosity $\mu_a = \mu$, we have four dimensional parameters $\mu, G, \rho, R$.
		By the Buckingham $\Pi$ Theorem, $\gamma$ is the only dimensionless parameter in this system.
		It can be interpreted as the number of oscillations a viscoelastic sphere undergoes before returning to the equilibrium.
		Incidentally, we can also consider the relaxation time scale of the viscoelastic solid, $\tau_\text{rel} = \mu / G$.
		Taking the ratio $\tau_\text{osc} / \tau_\text{rel}$ also yield the dimensionless parameter $\gamma$.

		\qquad We carry out a series of simulations (parameters documented in Table~\ref{table:str_params}) to demonstrate the scaling behavior in this system.
		In Fig.~\ref{fig:str_scaling}(a), we show solid elastic energy from various simulations normalized by its initial value $E_0$.
		Values of $G$, $R$, and $\mu$ are varied across simulations, but dimensionless parameter $\gamma$ is kept the same.
		The number of visible oscillations before energy is fully dissipated by the fluid is the same across these simulations.
		After rescaling time by $\tau_\text{diss}$, we also find that peaks and valleys of the elastic energy fall at similar places on the rescaled time axis.
		Furthermore, given the same $\gamma$, if the sphere radius is kept the same, which implies $\sqrt{G} / \mu$ is constant, elastic energy curves collapse onto the same one (simulation $A_1$ and $A_2$).

		\qquad In Fig.~\ref{fig:str_scaling}(b), we show the number of visible oscillations is qualitatively linearly proportional to $\gamma$.
		Simulation $B_1$, which has $\gamma = 10$, is compared to simulation $B_2$, $A_1$, and $B_3$, which have have $\gamma = 15.8, 20, 30$, respectively.
		The orange dashed line indicates approximately one oscillation period in simulation $B_1$ after initial release.
		Within the same rescaled time (in units of respective $\tau_\text{diss}$), we observe about 1.5, 2, and 3 times more periods in simulation  $B_2$, $A_1$, and $B_3$, respectively. The increase in the number of periods follows the increase in $\gamma$ approximately linearly.

		\qquad If the solid has a different density from fluid, the matter becomes more complex.
		Though the oscillation time scale $\tau_\text{osc}$ should only depend on $\rho_s$, the energy dissipation, influenced by dynamics in both the solid and the fluid phases, now depends on both $\rho_s$ and $\rho_f$.
		We demonstrate this effect in Fig.~\ref{fig:str_scaling}(c).
		In simulation $C_1$ and $C_2$, we have chosen to use $\rho_s$ in calculating $\gamma$ and $\tau_\text{diss}$.
		While the number of visible oscillations remains similar among simulations $A_1$, $C_1$, and $C_2$, which have identical $\gamma$,
		rescaling time with $\tau_\text{diss}$ no longer lead to the alignment of energy peaks and valleys on the rescaled time axis across simulations.
	\end{enumerate}

	\subsection{Settling}
		We provide \texttt{si\textunderscore movie\textunderscore 1.mov} {Settling of 150 ellipsoids, described in main text Fig.~3.}
	
	\subsection{Lid-driven Cavity} 
\begin{figure}
            \centering
                \includegraphics[width=\textwidth]{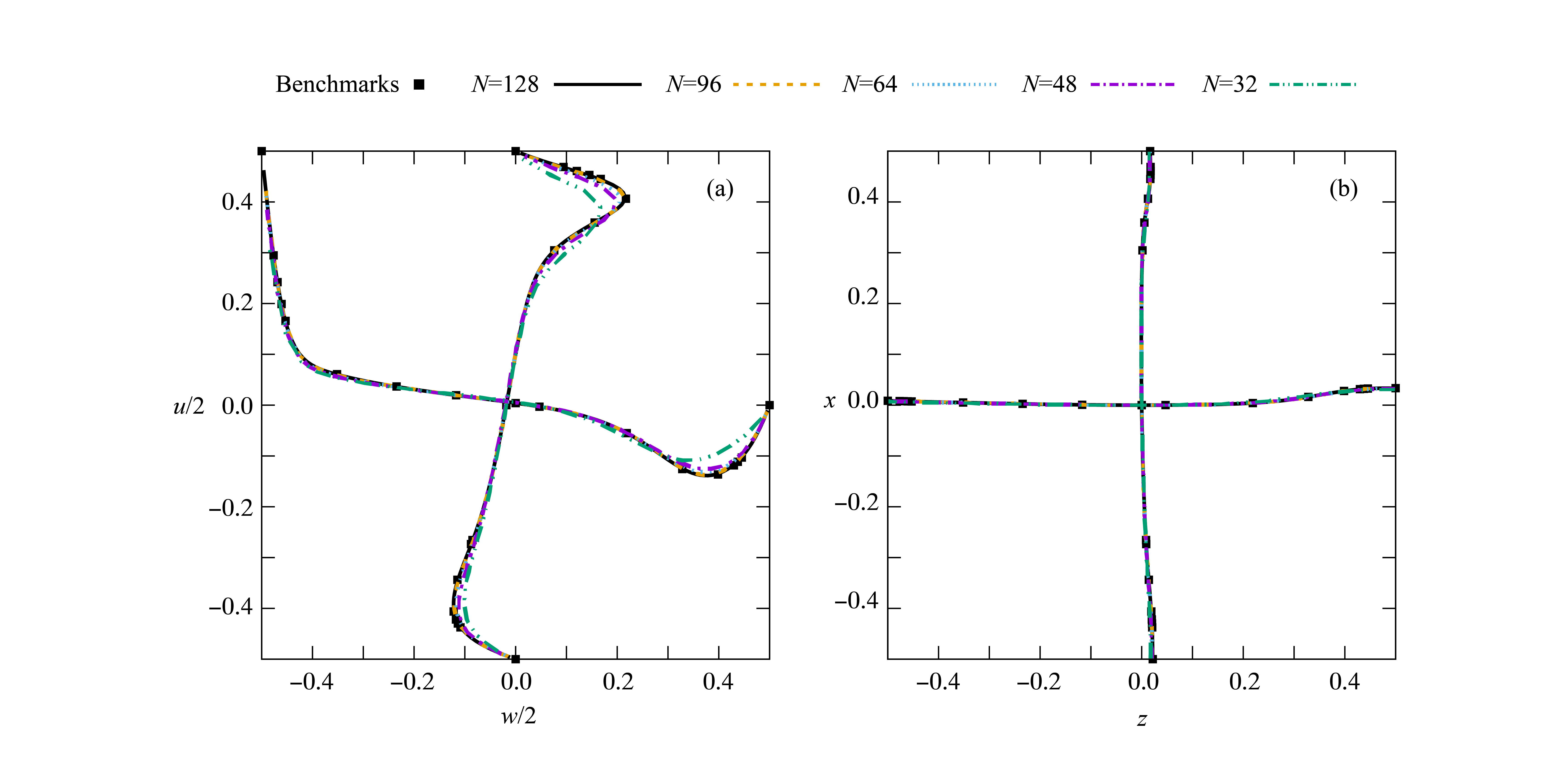}
                \caption{
                Comparing normal velocities and pressure profiles against benchmark results by Albensoeder and Kuhlmann \cite{albensoeder05} for cubic lid-driven cavity with Reynolds number $\Rey=1000$.
                Spatial coordinates are shifted by $0.5$ in both plots such that the domain is $[-0.5,0.5]\times[-0.5,0.5]$, and rotated to be aligned with figures in the work by Albensoeder and Kuhlmann \cite{albensoeder05}.
                (a) Normal velocities along the center lines in the central $xz$-plane are rescaled and plotted for five simulation
                resolutions, $N=32, 48, 64, 96, 128$. (b) Pressure profiles along the center lines in the central $xz$-plane are plotted. Pressure
                values are shifted by a constant such that it is kept at zero at position $(0.0, 0.0)$.}
                \label{fig:aspect1_profiles_re1000}
            \end{figure}

\begin{figure}
            \centering
                \includegraphics[width=\textwidth]{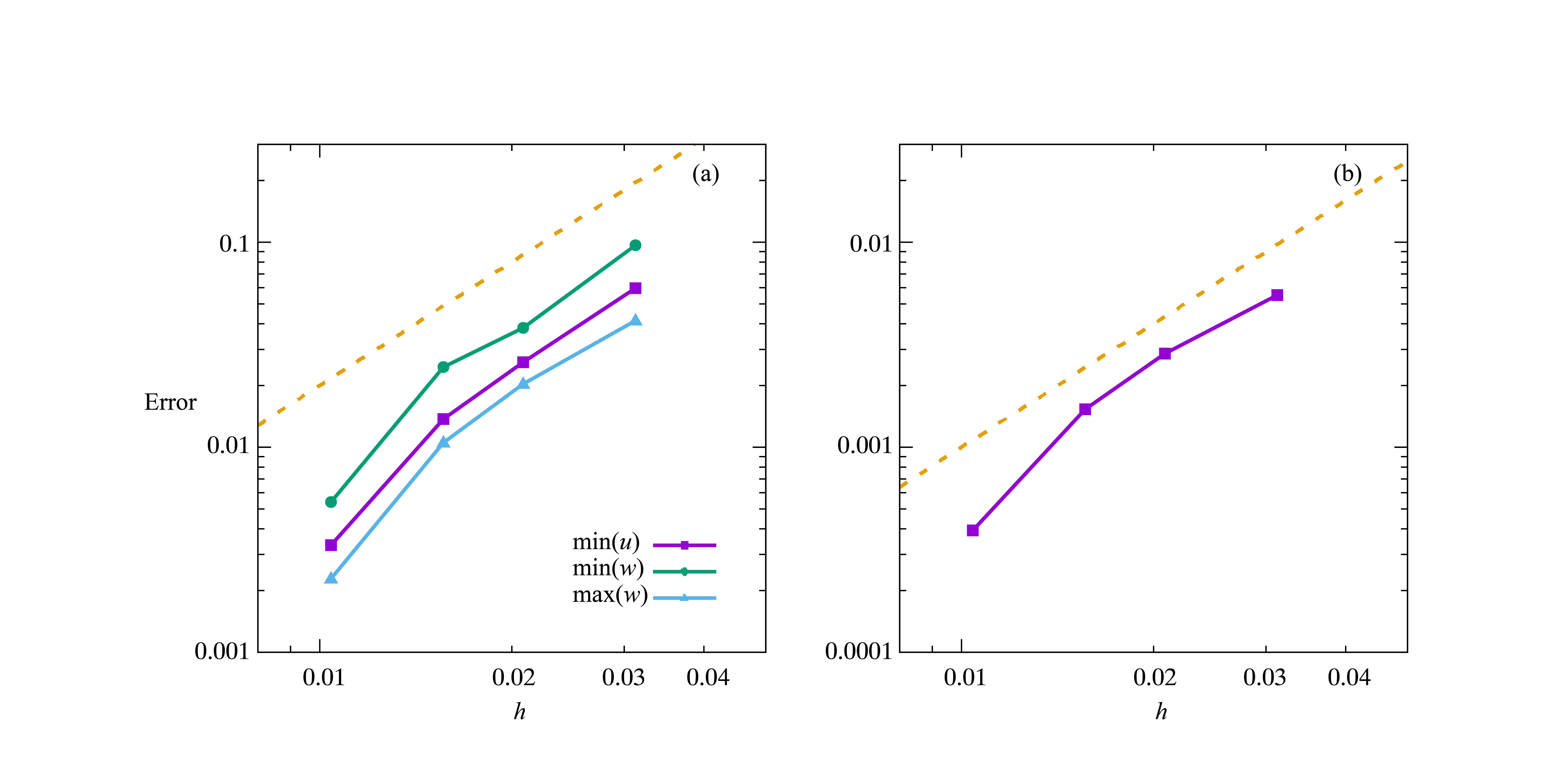}
                \caption{Error convergence in a cubic lid-driven cavity flow with $\Rey=1000$, using $N=128$ solution of as the reference solution. (a) Error in the extrema of normal velocities, $\min(u)$, $\min(w)$, and $\max(w)$, along center lines in the central $xz$-plane. (b) Error in the total kinetic energy after steady state has been reached.}\label{fig:aspect1_conv}
            \end{figure}

\begin{figure}
            \centering
                \includegraphics[width=0.75\textwidth]{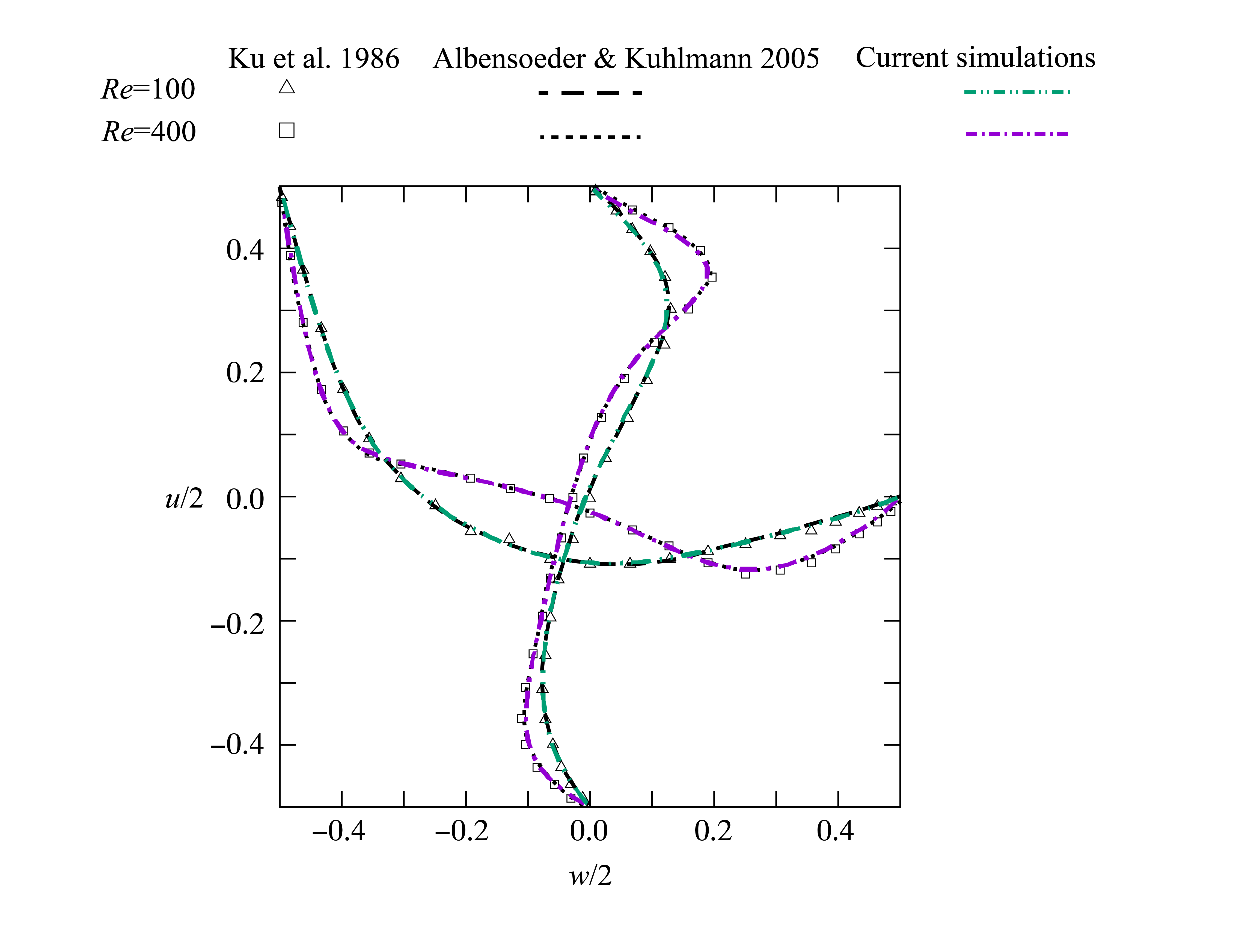}
                \caption{
                Comparing normal velocities against benchmark results by Albensoeder and Kuhlmann \cite{albensoeder05} for cubic lid-driven cavity with Reynolds number $\Rey=100, 400$.
                Spatial coordinates are shifted by $0.5$ in both plots such that the domain is $[-0.5,0.5]\times[-0.5,0.5]$, and rotated to be aligned with figures in the work by Albensoeder and Kuhlmann \cite{albensoeder05}.
                Normal velocities along the center lines in the central $xz$-plane are rescaled and plotted for resolution $N=96$.}
                \label{fig:aspect1_profiles_re400}
            \end{figure}

\begin{figure}
            \centering
                \includegraphics[width=\textwidth]{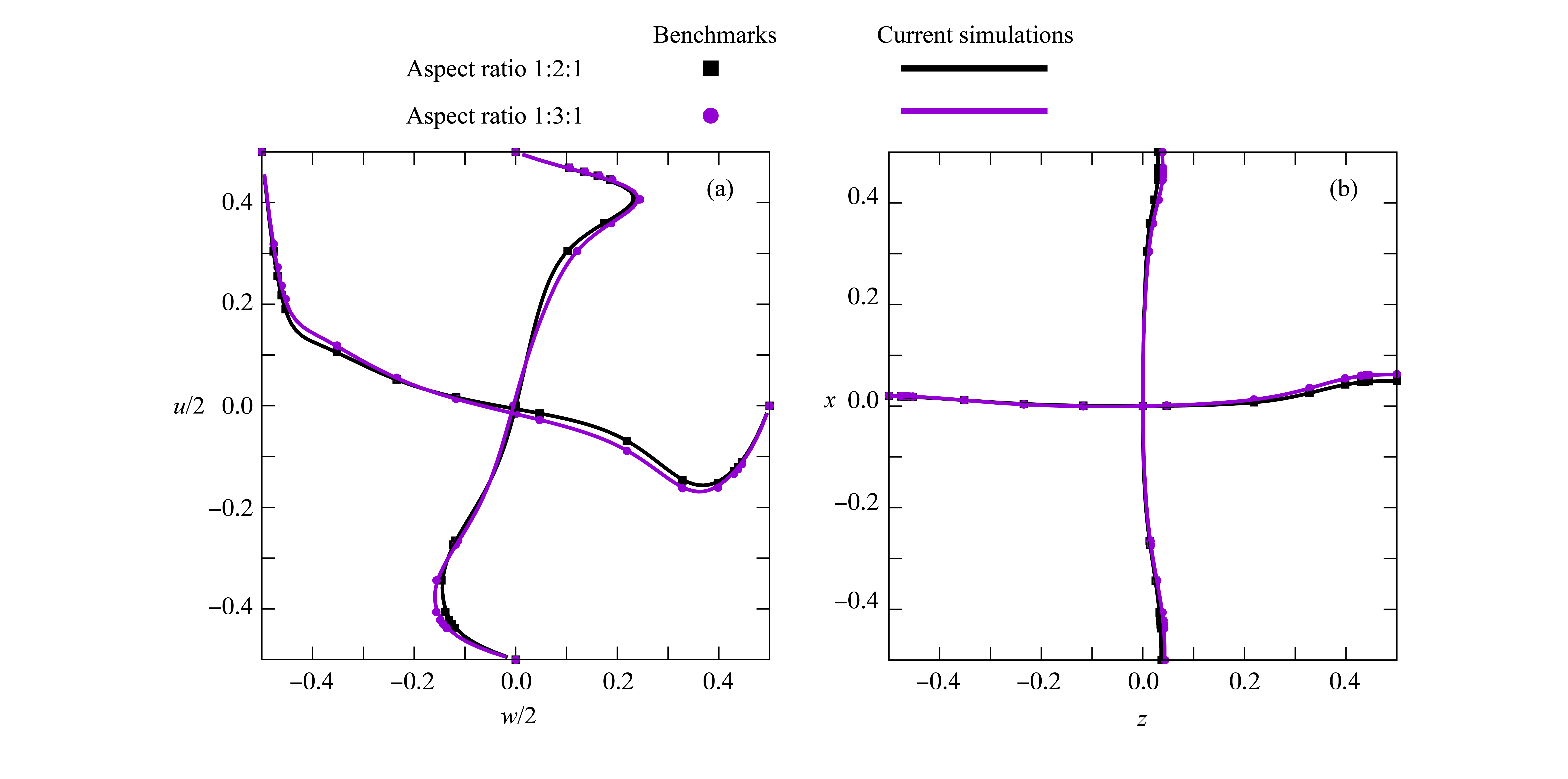}
                \caption{Comparing normal velocities and pressure profiles against benchmark results by Albensoeder and Kuhlmann \cite{albensoeder05} for lid-driven cavities with aspect ratios $1:2:1$ and $1:3:1$, with Reynolds number $\Rey=1000$.
                Spatial coordinates are shifted by $0.5$ in both plots such that the domain is $[-0.5,0.5]\times[-0.5,0.5]$, and rotated to be aligned with figures in the work by Albensoeder and Kuhlmann \cite{albensoeder05}.
                (a) Normal velocities along the center lines in the central $xz$-plane are rescaled and plotted for five simulation
                resolutions $N=96$ per unit simulation length. (b) Pressure profiles along the center lines in the central $xz$-plane are plotted. Pressure
                values are shifted by a constant such that it is kept at zero at position $(0.0, 0.0)$. }\label{fig:aspect23_profiles}
            \end{figure}

\begin{figure}
            \centering
                \includegraphics[width=\textwidth]{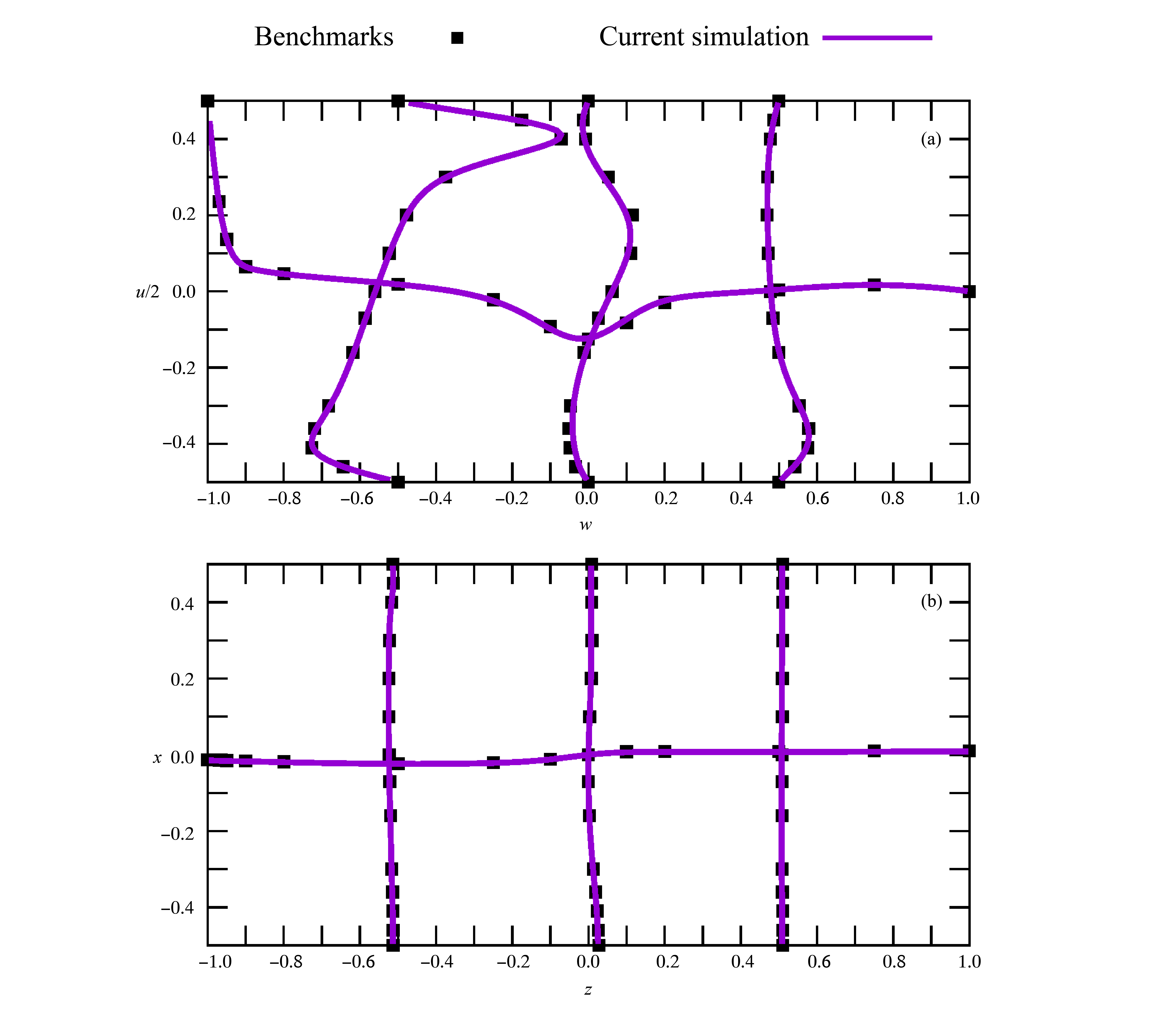}
                \caption{Comparing normal velocities and pressure profiles against benchmark results by Albensoeder and Kuhlmann \cite{albensoeder05} for lid-driven cavities with aspect ratio $1:1:2$, with Reynolds number $\Rey=1000$.
                Spatial coordinates are shifted in both plots such that the domain is $[-0.5,0.5]\times[-1.0,1.0]$, and rotated to be aligned with figures in the work by Albensoeder and Kuhlmann \cite{albensoeder05}.
                (a) Normal velocities along the center lines in the central $xz$-plane are rescaled and plotted for five simulation
                resolutions $N=96$ per unit simulation length. (b) Pressure profiles along the center lines in the central $xz$-plane are plotted. Pressure
                values are shifted by a constant such that it is kept at zero at position $(0.0, 0.0)$. }\label{fig:aspectz2_profiles}
            \end{figure}

\begin{figure}
            \centering
                \includegraphics[width=\textwidth]{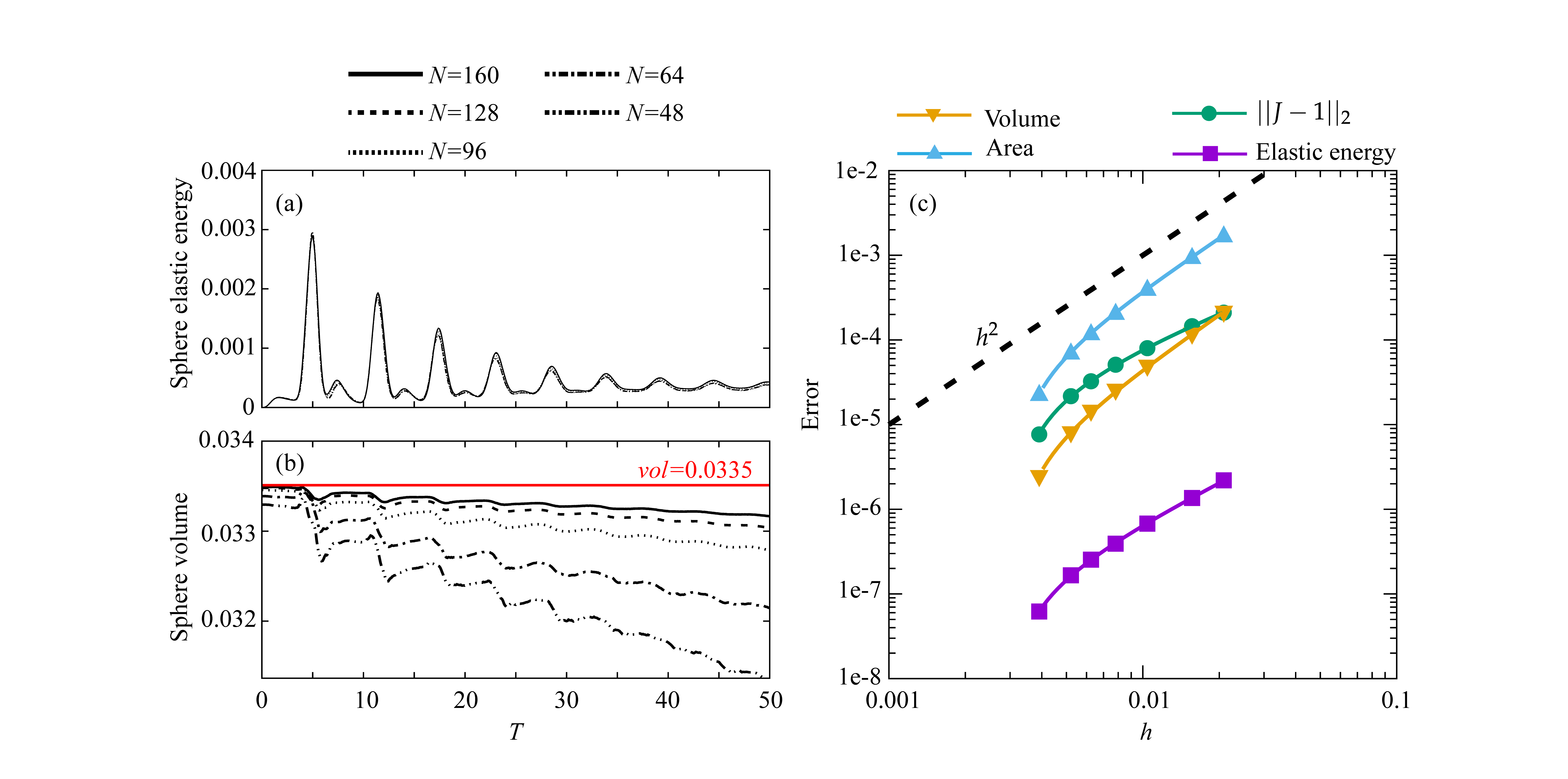}
                \caption{ (a) Sphere elastic energy and volume over time for a sphere with radius $0.2$ and shear modulus $G=0.1$ in a cubic lid-driven cavity.
                Reynolds number $\Rey=100$.
                 An incompressible neo-Hookean solid is modeled.
                 However,  $J \ne 1$ in our numerical scheme, especially near the fluid--solid interface.
                 Therefore, a correction factor is used in the strain energy density (Eq.~\eqref{eq:elas}).
                 The correction approaches zero as the grid is refined, since $J$ converges to 1.
                (b) Convergence of errors in surface area, volume, elastic energy, and $\|J-1\|_2$, where $J$ is the determinant of the deformation gradient.
                Errors are measured at $T=0.5$.
                The solid lines are fitted curves using Eq.~\eqref{eq:3params}.
                Convergence rates (with Richardson correction factor $\alpha$ in brackets) are $2.00 (0.98), 2.08 (1.0), 1.51 (0.92), \text{~and~} 1.16 (1.0)$ for surface area, volume, elastic energy, and $\|J-1\|_2$, respectively.
                }
                \label{fig:object_conv}
            \end{figure}

Here we provide results in comparing lid-driven cavity flow simulated with RMT3D without any solid objects with 3D benchmarks as a validation test for our Navier--Stokes solver code.
	In addition, we present convergence tests and centroid positions data of an immersed sphere in cubic lid-driven cavity flow.
	Convergence rates are computed using Eq.~\eqref{eq:3params} and reported in the figure caption when applicable.
	Finally, we provide additional simulation parameters used in Fig.~4 in the main text, and include movies mentioned there.
	\begin{enumerate}
		\item Comparison of normal velocities and pressure profiles in a cubic lid-driven cavity with Reynolds number $\Rey=1000$ against benchmark results is shown in Fig.~\ref{fig:aspect1_profiles_re1000}; convergence of normal velocity extrema and steady state kinetic energy is shown in Fig.~\ref{fig:aspect1_conv}.

		\item Comparison of normal velocities and pressure profiles in a cubic lid-driven cavity with Reynolds number $\Rey=100, 400$ against benchmark results is shown in Fig.~\ref{fig:aspect1_profiles_re400}.

        \item Comparisons of normal velocities and pressure profiles in a lid-driven cavity of various aspect ratios with $\Rey=1000$ against benchmark results are shown in Figs.~\ref{fig:aspect23_profiles} \& \ref{fig:aspectz2_profiles}.

		\item Convergence in surface area, volume, elastic energy,  and $\|J-1\|_2$, where $J$ is the determinant of the deformation gradient, for a sphere with $G=0.1$ in a cubic lid-driven cavity with $\Rey=100$ is shown in Fig.~\ref{fig:object_conv}.

		\item Additional parameters for simulations presented in Fig.~4 in the main text are documented in Table~\ref{table:lid_params}.

		\item \texttt{trajectories.txt} {trajectories.txt}{Trajectories of the centroid of a sphere with $G=0.1, 0.25, 0.5$ in a cubic lid-driven cavity with $Re=100$. Isotropic grid spacing $h=1/160$.}

		\item \texttt{si\textunderscore movie\textunderscore 2.mov} {A sphere of radius $0.2$ in unit cubic lid-driven cavity with $\Rey=100$ and solid shear modulus $G=0.03$, described in main text Fig.~4(a).}

		\item \texttt{si\textunderscore movie\textunderscore 3.mov}  {A sphere of radius $0.2$ in unit cubic lid-driven cavity with $\Rey=100$ and solid shear modulus $G=0.1$, described in main text Fig.~4(b)\&(c).}

		\item \texttt{si\textunderscore movie\textunderscore 4.mov} {A sphere of radius $0.2$ in unit cubic lid-driven cavity with $\Rey=100$ and solid shear modulus $G=0.25$, described in main text Fig.~4(d)\&(e).}

		\item \texttt{si\textunderscore movie\textunderscore 5.mov} {A sphere of radius $0.2$ in unit cubic lid-driven cavity with $\Rey=100$ and solid shear modulus $G=0.5$, described in main text Fig.~4(f)\&(g).}
	\end{enumerate}

    \subsection{Swimming}
        In this section, we expand on some of the contexts and results from the section in the main text on simulating swimming with the RMT.
        
        \subsubsection*{Amplitude parameter} The amplitude parameter $W = B/3GIk^2$ is derived from scaling arguments and linear bending theory.
        Here, we explicitly list the relationships and scalings between bending moment amplitude $B$, displacement scaling constant $W$, and the vertical linear stress density $\sigma_0$.
        
        In Euler beam theory, the stress density is related to the bending moment across a cross-section as $\sigma_0 \sim B/I$, where $I$ is the area moment of inertia of the cross-section.
        There is a similar relationship to the beam curvature $\kappa \sim B/3GI$, where $3GI$ is the beam bending modulus for a shear modulus $G$.
        Given a vertical displacement scale $W$ and length scale $k^{-1}$, the curvature also satisfies $\kappa \sim Wk^2$ for $Wk\ll 1$. Thus, $W \sim B/3GIk^2$ and $\sigma_0 \sim 3WGk^2$, motivating the definitions in the main text.
    
        \subsubsection*{Swimming statistics} For each simulation, we calculate a swim speed $U$, active power $P$, drag coefficient $C$, and swimming efficiency $e$. Here, we detail the form of each and describe its calculation.
        
        We let $\left<\psi\right>$ describe the time-average and $\hat{\psi}$ the oscillation magnitude of a time-dependent value $\psi(t)$. To calculate these quantities, time traces $\psi(t)$ are fit to a function
        \begin{equation}
            f(t) = c_0 + c_1 t + c_2 t^2 + \sum_{j=1}^2 \left[A_j \cos(2\pi j t) + B_j \sin(2 \pi j t)\right].
        \end{equation}
        Then $\left<\psi\right> = c_0$, $\hat{\psi} = \sqrt{A_1^2+B_1^2}.$
        
        \begin{itemize}
            \item The swim speed $U = \left<u^{(c)}_x\right>$, where $\vu^{(c)}$ is centroid velocity and the subscript $x$ denotes the direction of swimming
            \item The power $P = \left<-\int_{\Omega_s} \vsigma^{(a)}:\nabla \vu dV\right>$
            where $\vsigma^{(a)}$ is the active part of the solid stress
            \item The drag coefficient $C$ is computed for each combination of swimmer body radius $R$ and length $L$.
            A simulation is conducted for each body shape with no active stress subject to a density ratio $\rho_s / \rho_f = 2$ and gravitational acceleration $g = 1/10$ in the swimming direction.
            The centroid velocity $u_x^{(c)}(t)$ is fit to the function
            \begin{equation}
                g(t) = U_f + (U_0-U_f) e^{-\lambda t},
            \end{equation}
            for an initial and final velocity $U_0$ and $U_f$ and a rate $\lambda$ representing the speed with which the object approaches terminal velocity. The drag coefficient is then calculated as $C = (\rho_s - \rho_f) V_s g / U_f$, where $V_s$ is the swimmer volume.
            \item The swimming efficiency is typically defined $e = F U/P$, where $F$ is the force required to tow the swimmer at velocity $U$.
            In the main text, we substitute an approximate tow force $\bar F = CU$ representing a linear estimate.
        \end{itemize}
    
        \subsubsection*{Reynolds numbers} We describe two Reynolds numbers: one corresponding to the steady flow induced by time-averaged motion of the object through the fluid, and another describing the oscillatory flow driven by the object's cyclic deformation. Here we describe the form of both and report the calculated values.
        
        \begin{itemize}
            \item The steady Reynolds number $\Rey_s = \rho U R_h/\mu$ uses the calculated swim speed as a velocity scaling and the swimmer head radius as a length scale.
            \item The oscillatory Reynolds number $\Rey_o = \rho \hat{u}^{(c)}_z R / \mu$ uses $\hat{u}^{(c)}_z$---the magnitude of the oscillating vertical velocity---as a velocity scale and the swimmer radius as a length scale.
        \end{itemize}
        
\begin{figure}
                \centering
                \includegraphics[width=0.7\textwidth]{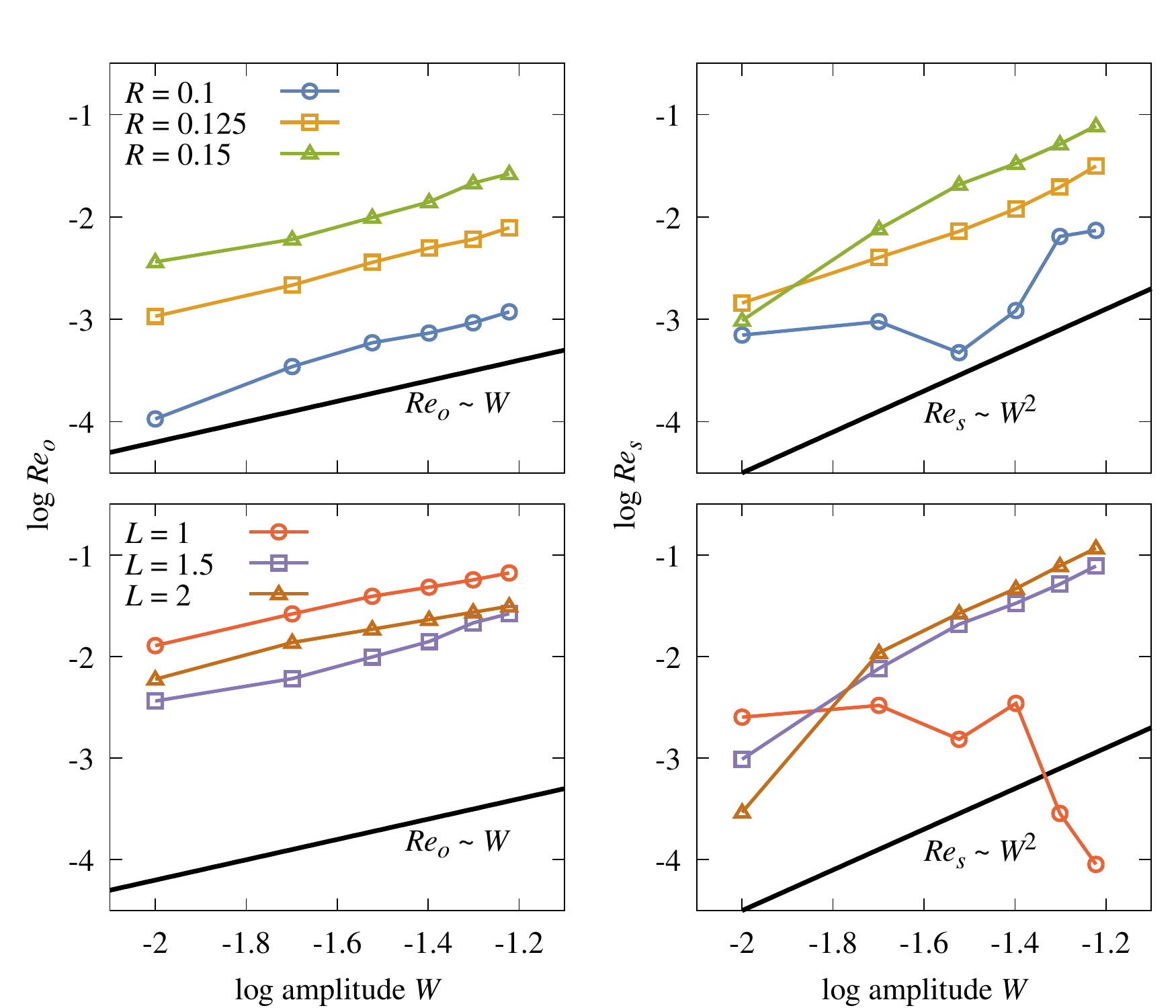}
                \caption{Reynolds numbers describing the oscillatory ($\Rey_o$, left column) and steady ($\Rey_s$, right column) flows about the swimmer.
                Top row: Swimmer body length is kept at $L=1.5$ while its body radius $R$ varies.
                Bottom row: Swimmer body radius is kept at $R=0.15$ while its body length $L$ varies.
                Other simulation parameters are reported in Fig.~5 caption in the main text.
                }
                \label{fig:renums}
            \end{figure}

        For each simulation reported in Fig.~5 in the main text, the values of the two Reynolds numbers are calculated and documented in Fig.~\ref{fig:renums}.

          \subsubsection*{Movie} Finally, we provide \texttt{si\textunderscore movie\textunderscore 6.mov}
        {Swimmers with various body geometries, actuated by active stress, described in main text Fig.~5.}

\newpage

            \begin{table}\centering
                \caption{Main dimensionless simulation parameters in the RMT.
                In addition, Dirichlet and Neumann boundary conditions can be specified for each face of the rectangular simulation domain.}
                \label{table:pars}

                \begin{tabular} { c || l }
                    \hline
                    $\rho_f$ 	& fluid density \\\hline
                    $\mu$       	& fluid viscosity \\\hline
                    $\rho_s$ 	& solid density \\\hline
                    $G$      		& solid shear modulus \\\hline
                    $\mu_a$ 	& solid artificial viscosity \\\hline
                    $\gamma_t$  & blur zone viscosity constant \\\hline
                    $\epsilon$ 	& half blur zone width\\\hline
                    $L_x, L_y, L_z$ & simulation domain side lengths \\\hline
                \end{tabular}
            \end{table}

   	        \begin{table}\centering
                \caption{Dimensionless simulation parameters for pre-stretched sphere scaling test. Fluid density $\rho_f$ is always kept at unity. Solid artificial viscosity $\mu_a$ is identical to fluid viscosity $\mu$.}
                \label{table:str_params}
                \begin{tabular} { c || c | c | c | c || c | c}
                    \hline
                    Case       	& $\mu$ & $\rho_s/\rho_f$ & $R$ & $G$  & $\gamma$  & $\tau_\text{diss}$ \\ \hline
                    $A_1$	& 0.005     &	1	& 0.2   & 0.25  & 20.0	& 8 \\ \hline
                    $A_2$	& 0.01      &	1	& 0.2   & 1     & 20.0 	& 4 \\ \hline
                    $A_3$	& 0.005     &	1	& 0.1   &  1    & 20.0 	& 2 \\ \hline
                    $A_4$	& 0.0075    &	1	& 0.3   & 0.25  & 20.0	& 12 \\ \hline
                    \hline
                    $B_1$	& 0.005 &   1   & 0.1   & 0.25 & 10.0  & 2 \\ \hline
                    $B_2$	& 0.02  &	1	& 0.2   & 2.5  & 15.8 	& 2 \\ \hline
                    $B_3$	& 0.005 &	1	& 0.3   & 0.25 & 30.0 	& 18 \\ \hline
                    \hline
                    $C_1$	& 0.005 &   0.5625  & $\tfrac{4}{15}$   & 0.25  & 20.0  & 8 \\ \hline
                    $C_2$	& 0.005  &	2.25    & $\tfrac{2}{15}$   & 0.25 	& 20.0 	& 8 \\ \hline
                 \end{tabular}
            \end{table}

   	        \begin{table}\centering
                \caption{Additional parameters simulations presented in Fig.~4 in the main text. Number of grid cells $N$ and blur zone viscosity multiplier $\gamma_t$ are reported for each case in subfigure Fig.~4(a), (b), (d), (f), (h), and (j). Isotropic grid spacing $h=1/N$.}
                \label{table:lid_params}
                \begin{tabular} { c || c | c }
                    \hline
                    Case       	& $N$ & $\gamma_t$ \\ \hline
                    $(a)$	& 128 &	0	\\ \hline
                    \hline
                    $(b)$	& 48        &	2	\\ \hline
                    $(b)$	& 64        &	4	\\ \hline
                    $(b)$	& 96        &	2	 \\ \hline
                    $(b)$	& 128       &	0	 \\ \hline
                    $(b)$	& 160       &	0	 \\ \hline
                    \hline
                    $(d)$	& 48        &	0	\\ \hline
                    $(d)$	& 64        &	1	\\ \hline
                    $(d)$	& 96        &	0	 \\ \hline
                    $(d)$	& 128       &	0	 \\ \hline
                    $(d)$	& 160       &	0	 \\ \hline
                    \hline
                    $(f)$	& 48        &	0	\\ \hline
                    $(f)$	& 64        &	0	\\ \hline
                    $(f)$	& 96        &	0	 \\ \hline
                    $(f)$	& 128       &	0	 \\ \hline
                    $(f)$	& 160       &	0	 \\ \hline
                    \hline
                    $(h)$	& 128       &	0	 \\ \hline
                    \hline
                    $(j)$	& 64        &	0	 \\ \hline
                    \hline
                 \end{tabular}
            \end{table}

\FloatBarrier
\renewcommand\refname{Supplmentary Information References}
\bibliographystyle{unsrt}
\bibliography{si-references}